\pdfoutput=0
\documentclass[twocolumn]{emulateapj}

\usepackage{graphicx}
\usepackage{natbib}
\usepackage{multirow}
\usepackage{amsmath}
\usepackage{amssymb}
\usepackage{longtable}

\newcommand{\bc}{\begin{center}}
\newcommand{\ec}{\end{center}}

\newcommand{\kms}{km~s$^{-1}$ }

\newcommand{\hco}{HCO$^{+}$}
\newcommand{\hcoiso}{H$^{13}$CO$^{+}$}
\newcommand{\cm}{cm$^{-3}$}

\begin{document}

\slugcomment{}

\title{The Excitation of HCN and \hco\, in the Galactic Center Circumnuclear Disk}

\author{E.A.C. Mills\altaffilmark{1}}
\affil{National Radio Astronomy Observatory\altaffilmark{2},}
\affil{Department of Physics and Astronomy, University of California}
\affil{Physics and Astronomy Building, 430 Portola Plaza, Box 951547 Los Angeles, CA 90095-1547}
\email{bmills@aoc.nrao.edu}

\author{R. G\"{u}sten}
\affil{Max Planck Institut f\"{u}r Radioastronomie}
\affil{ Auf Dem Huegel 69, 53121 Bonn, Germany }

\author{M.A. Requena-Torres}
\affil{Max Planck Institut f\"{u}r Radioastronomie} 
\affil{Auf Dem Huegel 69, 53121 Bonn, Germany }

\author{M.R. Morris}
\affil{Department of Physics and Astronomy, University of California}
\affil{Physics and Astronomy Building, 430 Portola Plaza, Box 951547 Los Angeles, CA 90095-1547}
\altaffiltext{1}{E.A.C. Mills is a Jansky Fellow of the National Radio Astronomy Observatory.}
\altaffiltext{2}{The National Radio Astronomy Observatory is a facility of the National Science Foundation operated under cooperative agreement by Associated Universities, Inc.}

\begin{abstract}

We present new observations of HCN and \hco\, in the circumnuclear disk (CND) of the Galaxy, obtained with the APEX telescope.
We have mapped emission in rotational lines of HCN J = 3--2, 4--3, and 8--7, as well as \hco\, J = 3--2, 4--3, and 9--8. We also present spectra of H$^{13}$CN J = 3--2 and 4--3, and \hcoiso\, J = 3--2 and 4--3 toward four positions in the CND. Using the intensities of all of these lines, we present an excitation analysis for each molecule using the non-LTE radiative transfer code RADEX. The HCN line intensities toward the northern emission peak of the CND yield log densities (cm$^{-3}$) of 5.6$^{+0.6}_{-0.6}$, consistent with those measured with \hco\, as well as with densities recently reported for this region from an excitation analysis of highly-excited lines of CO. These densities are too low for the gas to be tidally stable. The HCN line intensities toward the CND's southern emission peak yield log densities of 6.5$^{+0.5}_{-0.7}$, higher than densities determined for this part of the CND with CO (although the densities measured with \hco, log [n] = 5.6$^{+0.2}_{-0.2}$, are more consistent with the CO-derived densities).  We investigate whether the higher densities we infer from HCN are affected by mid-infrared radiative excitation of this molecule through its 14 $\mu$m rovibrational transitions. We find that radiative excitation is important for at least one clump in the CND,  where we additionally detect the J = 4--3, $v_2=1$ vibrationally-excited transition of HCN, which is excited by dust temperatures of $\gtrsim$ 125-150 K. If this hot dust is present elsewhere in the CND, it could lower our inferred densities, potentially bringing the HCN-derived densities for the Southern part of the CND into agreement with those measured using \hco\, and CO. Additional sensitive, high-resolution submillimeter observations, as well as mid-infrared observations, would be useful to assess the importance of the radiative excitation of HCN in this environment. 
\end{abstract}
\keywords{Galaxy: center- radiolines: ISM-ISM: clouds- ISM: molecules }

\section{Introduction}
The Circumnuclear Disk (CND) is a ring of gas and dust around the central supermassive black hole (SMBH), with an inner radius of $\sim$ 1.5 pc and an inclination of  70 degrees from the Galactic plane \citep{Gusten87,Jackson93}. The CND has been studied for decades in a range of molecular transitions;  \cite{AB11} provide a comprehensive list of molecular lines studied in the CND to date. The molecular gas temperatures determined from observations of CO range from 50 - 400 K \citep{Harris85, Lugten87, Bradford05, Oka11}. Atomic gas temperatures measured at the inner edge of the CND are comparable, ranging from 200 to 350 K  \citep{Genzel85, Jackson93}. Measured dust temperatures are much lower, ranging from 20 to 90K \citep[][Lau et al. 2013]{Becklin82, Mezger89, Etx11, Molinari11}. Kinematic studies of the CND \citep{Jackson93,Wright01,Martin12,Liu12} show that the motion of the bulk of its gas is consistent with orbiting filaments in several planes. 

The CND is the closest large molecular structure to the SMBH  \citep[mass $\sim 4.5 \times 10^6$ M$_{\odot}$;][]{Ghez08, Gillessen09,Genzel10}, and thus should be subject to strong tidal shearing forces. If the CND is to be stable against tidal shearing, it must have an extremely high density: the fluid Roche approximation for gas at a radius of 2 pc from the SMBH yields a limiting density of 8$\times10^7$ \cm. Observations of molecular clumps in the CND with lines of HCN and \hco\, have led to virial density estimates for individual clumps that range from $10^7$ to $10^8$ cm$^{-3}$ \citep{Shukla04,Chris05, MCHH09}. However, the high densities derived in this way are in disagreement with lower densities recently inferred from the dust emission characteristics \citep{Etx11} and from an excitation analysis using the CO molecule \citep[][hereafter RT12]{RT12}. 

Both \citeauthor{Chris05} and \citeauthor{MCHH09} also estimate a total mass for the CND ($\sim 10^6$ M$_{\odot}$), by assuming virialization, which greatly exceeds previous mass estimates of a few $\times 10^4$ M$_{\odot}$, based on CO excitation and dust emission \citep{Harris85, Genzel85, Lugten87, Mezger89}.  More recent estimates of the mass from the FIR and submillimeter dust emission also favor a lower mass:  \cite{Etx11} estimate a mass of $5 \times 10^4$ M$_{\odot}$, and RT12 estimate a mass of  10$^3$ - 10$^4$ M$_{\odot}$. Lau et al. (2013) measure an even lower mass, $\sim$ 600 M$_{\odot}$, but just for the relatively warm ``circumnuclear ring" of material at the inner edge of the CND. Recent observations then suggest that the high values derived for the CND mass and density are due to an (invalid) assumption of virialization for the bulk of the gas. However, due to the nature of the tracers used in each analysis (HCN and \hco, versus the lower-density tracer CO and dust), it is not clear whether recent observations rule out the existence of any virialized clumps in the CND.  Verifying whether there is any CND gas for which the assumption of virialization is valid is important for constraining the mass of the molecular gas reservoir in the central two parsecs of the galaxy, and for determining whether the gas densities are consistent with recent suggestions of star formation in the CND \citep{YZ08}, given the strong tidal forces which are present in this region.  \vspace{0.3cm}

In order to investigate whether higher-density molecular tracers indicate the presence of virialized gas in the CND, we have conducted an excitation analysis using the $J$ = 3--2, 4--3 and 8--7 transitions of HCN and the $J$ = 3--2, 4--3, and 9--8 transitions of \hco\, from single-dish observations with the Atacama Pathfinder Experiment telescope (APEX\footnote{This publication is based on data acquired with the Atacama
Pathfinder EXperiment. APEX is a collaboration between the Max-Planck-Institut f\"{u}r Radioastronomie, the European Southern
Observatory, and the Onsala Space Observatory.}). These lines should be particularly sensitive to the presence of dense gas, having critical densities of $10^6 - 10^9$ \cm, and spanning a range of energies from 25 K to 190 K above the ground state. Single-dish observations also have the advantage that they are not subject to the filtering out of spatially-extended flux, which may have affected previous interferometric studies.  This is the first excitation analysis of the CND using these dense gas tracers, and the first detection of the highly excited HCN $J$ = 8--7 and \hco\, $J$ = 9--8 lines in the CND. In Section \ref{Obs}, we describe the APEX observations. In Section \ref{maps}, we present velocity-integrated maps and spectra of all of the observed transitions. In Sections \ref{Ana} and  \ref{Res} we discuss the excitation analysis and our derived constraints on the temperature and density of the CND.  Finally, we conclude in Section \ref{Dis} with a discussion of the excitation mechanisms for HCN in the CND, and an assessment of the validity of the assumption that all CND gas clumps are virialized. We also comment on whether the chemistry in the CND is likely to be affected by X-rays or cosmic rays. 

\section{Observations and Calibration}
\label{Obs}

All of the data used for the analysis in this paper were obtained with the APEX telescope \citep{Gusten06}, a single 12 meter modified ALMA prototype antenna located at an elevation of 5106 m on the Chajnantor plain in Chile. Data were obtained over a multiple-day run in July 2010. Conditions for the run were excellent, with a precipitable water vapor overburden of less than 0.5 mm. 
Additional observations of  H$^{13}$CN $J$ = 4--3 and \hcoiso\, $J$ = 4--3 were obtained in November 2010 toward the northern emission peak of the CND. Observations of H$^{13}$CN $J$ = 3--2, HC$^{15}$N $J$ = 3--2, H$^{13}$CN $J$ = 4--3 and HC$^{15}$N $J$ = 4--3  were obtained in April 2012 toward the southern emission peak of the CND and an additional pointing toward the Southwest component of this peak. Properties of all of the observed transitions are listed in Table \ref{lines}. 

\subsection{260 GHz observations}

Using the APEX1 facility receivers, we simultaneously observed the $J$ = 3--2 transition of HCN and the $J$ = 3--2 transition of \hco\, over a rectangular field covering the CND. Our maps were centered on the position of Sgr A*, at RA=17h 45m 39.92s, Dec=-29$\degr$00$'$28.1$''$ (J2000). In addition, we made pointed observations toward three positions in the CND in the $J$ = 3--2 transitions of the isotopologues  H$^{13}$CN and \hcoiso. The GILDAS\footnotemark[2] \footnotetext[2]{http://www.iram.fr/IRAMFR/GILDAS} software CLASS90 was used to reduce the calibrated data. We fit and removed a first-order baseline from all of the spectra, and boxcar-smoothed the spectra to a resolution of 5.2 km s$^{-1}$. A correction for a main-beam efficiency of 0.71 for the 2010 data and 0.75 for the 2012 data was also applied. The data were then gridded onto rectangular maps of $\sim 5.8 \times 2.3'$ extent \citep[13 $\times$ 7 parsecs at the assumed 8.4 kpc distance of the Galactic center;][]{Ghez08,Gillessen09}, with a pixel size of $11.8''$. The spatial resolution of the maps varies slightly for each line, but at these frequencies is $\sim$24\arcsec. The estimated calibration uncertainty of these data is 10\%. 

\subsection{350 GHz observations}

Using the FLASH receiver \citep{Heyminck06}, we also simultaneously mapped the $J$ = 4--3 lines of HCN and \hco\, over a field centered on Sgr A*. We additionally made pointed observations toward three positions in the CND in the $J$ = 4--3 transitions of the isotopologues  H$^{13}$CN and \hcoiso. All of these data were processed in the same manner as the 260 GHz observations, with the main-beam efficiency at the frequency of this line being 0.67 for the 2010 data and 0.73 for the 2012 data. The data were then gridded onto rectangular maps of $\sim 4.2' \times 2.3'$, with a pixel size of $8.9''$. The angular resolution of the maps at these frequencies is $\sim$18\arcsec. The estimated calibration uncertainty of these data is 10\%.

\subsection{700-800 GHz observations with CHAMP$^+$}
Using the CHAMP$^+$ heterodyne array receiver \citep{Gusten08, Kasemann06}, we simultaneously observed the HCN $J$ = 8--7 transition and the \hco\, $J$ = 9--8 (The \hco\, $J$ = 8--7 transition at 713 GHz is not observable from the ground, due to low atmospheric transmission at that frequency). Emission from both of these lines was mapped around two positions in the CND, toward the northern and southern emission peaks. These data were processed in the same manner as the 260 and 350 GHz observations. The HCN 8--7 data were then gridded onto rectangular maps of $\sim 1.5' \times 1.5'$ covering the southern emission peak of the CND and $1.3' \times 0.7'$ covering the northern emission peak, with pixel sizes of $5''$. The angular resolution for HCN 8--7 is 9\arcsec\,. The \hco\, 9--8 data were also gridded onto rectangular maps of $\sim 1.2' \times 1.2'$ (for the southern emission peak) and $1.0' \times 0.5'$ (for the northern emission peak), with pixel sizes of $5''$. A correction for a main-beam efficiency of 0.40 was applied, with the estimated calibration uncertainty for both lines being 20\%. 

\section {Results}
\label{maps}

\begin{figure*}[tbh]
\includegraphics{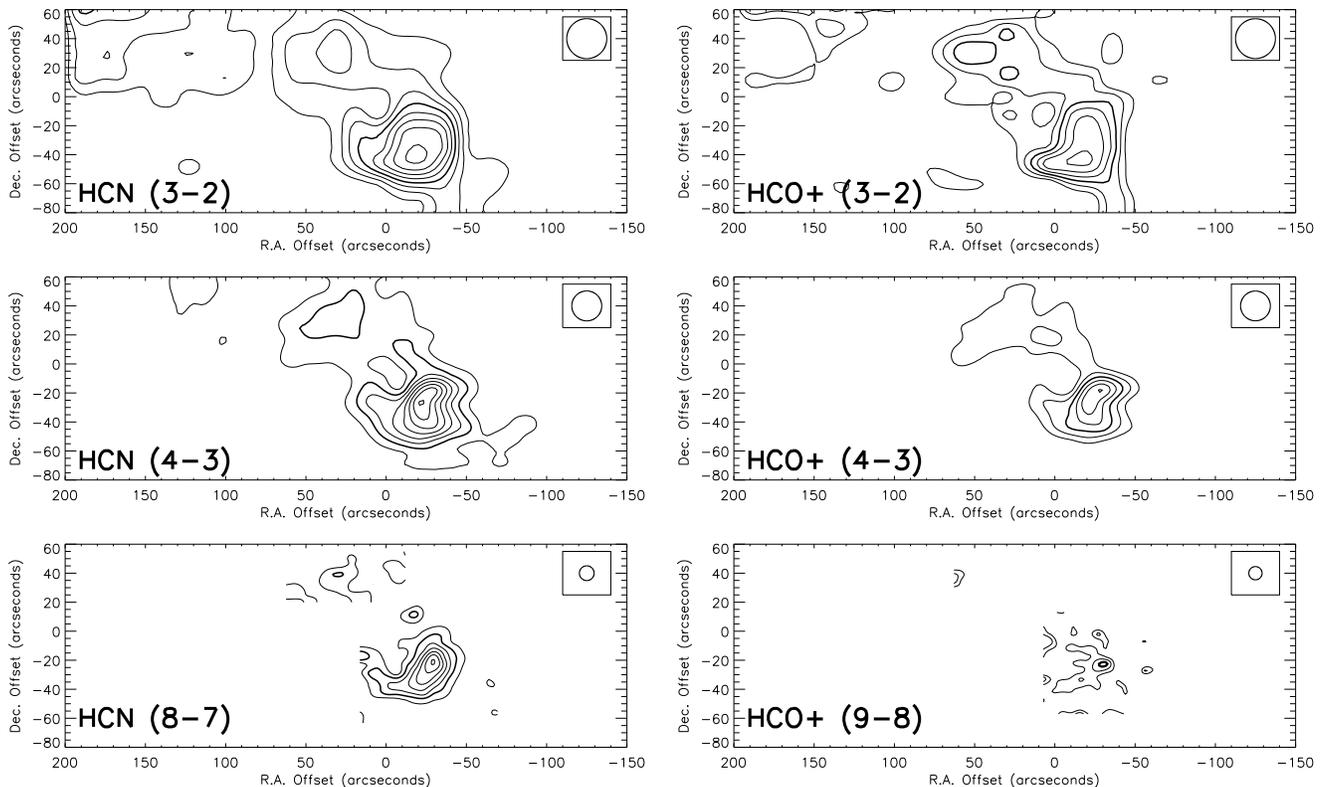}
\caption{  Contour maps of spatially unconvolved main beam brightness temperature integrated from -200 to 200 \kms\,\hspace{-0.1cm} for the observed lines of HCN and \hco.  Contours are linearly spaced. Top Left: HCN 3--2, contours from 750 to 2650 K \kms\hspace{-0.1cm}, spaced by 268 K \kms\hspace{-0.1cm}. Top Right: \hco\, 3--2, contours from 575 to 1400 K \kms\hspace{-0.1cm}, spaced by 166 K \kms\hspace{-0.1cm}. Middle Right: HCN 4--3, contours from 600 to 2620 K \kms\hspace{-0.1cm}, spaced by 288 K \kms\hspace{-0.1cm}. Middle Left: \hco\, 4--3, contours from 575 to 1400 K \kms\hspace{-0.1cm}, spaced by 169 K \kms\hspace{-0.1cm}. Bottom Left: HCN 8--7, contours from 150 to 810 K \kms\hspace{-0.1cm}, spaced by 110 K \kms\hspace{-0.1cm}. Bottom Right: \hco\, 9--8, contours from 110 to 250 K \kms\hspace{-0.1cm}, spaced by 67 K \kms\hspace{-0.1cm}. The beam sizes are given in Table \ref{lines}. }
\label{contour}
\end{figure*}

	Contour maps of HCN and \hco\, integrated intensity are shown in Figure \ref{contour}. For all the mapped lines, the strongest emission is seen toward the southern emission peak of the CND, consistent with the HCN 4--3 and CS 7--6 maps of \cite{MCHH09}. Emission from M-0.02-0.07, the 50 \kms\, cloud, can be seen in the northeast corner of the HCN and \hco\,  3--2 and 4--3 maps.  The HCN and \hco\, maps exhibit largely the same morphology, though the \hco\, intensity in a given transition is weaker than HCN in the same transition by a factor of $\sim$ 1.5-2. 

\begin{figure*}[tbh]
\hspace{3cm}
\includegraphics{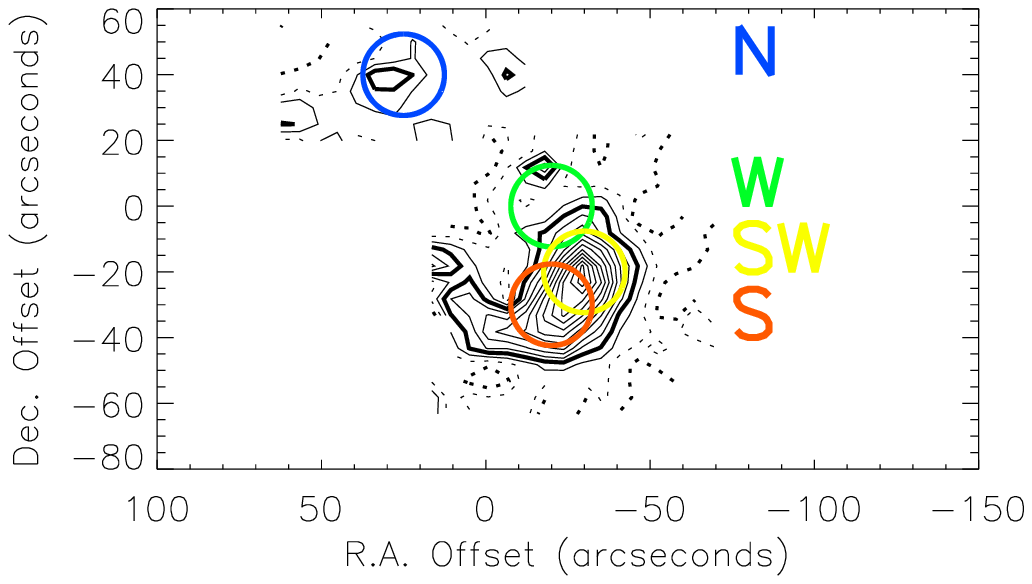}
\vspace{1.3cm}
\caption{  Contour map of unconvolved HCN 8--7 emission in the CND integrated over a velocity range from -200 \kms\, to 200 \kms\hspace{-0.1cm}. Contours are linearly spaced, from 0 to 810 K \kms\hspace{-0.1cm}. The circles are the size of the H$^{13}$CN 3--2 beam and show positions where spectra of the $^{13}$C isotopologues were obtained. }
\label{87_contour}
\end{figure*}

	\subsection{Pointed isotopologue observations}
		Pointed observations of the $^{13}$C isotopologues of HCN  were made toward four positions: the southern emission peak (RA, Dec offset\footnotemark[3]\footnotetext[3]{All offset positions are given in arcseconds with respect to the position of Sgr A*} = -20\arcsec, -30\arcsec), the southwest portion of the southern emission peak (-30\arcsec, -20\arcsec), the northern emission peak (+25\arcsec, +40\arcsec), and the western edge of the CND  (-20\arcsec, +0\arcsec) . Observations of \hcoiso\, were made toward three of these positions (North, South, and West). The locations of all four pointings are shown in Figure \ref{87_contour}. 

\begin{figure*}[tbh]
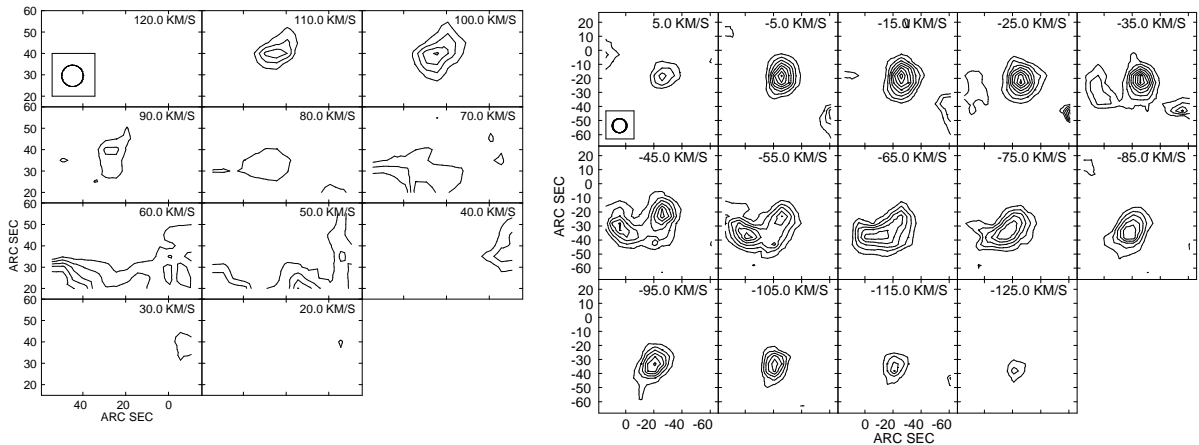

\includegraphics[scale=0.3, angle=270]{3-KNTR87N.PS}
\includegraphics[scale=0.35, angle=270]{3-KNTR87S.PS}
\caption{Channel maps of unconvolved HCN 8--7 emission toward the northern (Left) and southern (Right) emission peaks of the CND. Contours are linearly spaced by 0.21 K, from 0.43 to 2.13 K. Data are binned into 5 \kms\,\hspace{-0.1cm} channels, with every other channel pictured here.} 
\label{87_north}
\end{figure*}

		These pointings cover four (or possibly five) distinct features in position and velocity space (listed in Table \ref{Clumps}), the properties of which will be the focus of this paper. We identify these features in our HCN 8--7 cubes, as they have the highest spatial resolution. Channel maps of these cubes toward the northern and southern emission peaks of the CND are shown in Figure \ref{87_north}. The locations of the HCN 8--7 peak emission deviate slightly from the positions of the pointed isotopologue observations, but all features fall in the beam of the corresponding isotopologue pointings. Compared to H$^{13}$CN, the \hcoiso\, intensities for the same $J$-transition are typically a factor of 2.5-5 times weaker. The relative intensities of HCN and \hco\, are discussed further in Section \ref{env}.

\begin{figure*}[tbh]
\hspace{2.2cm}
\includegraphics[scale=0.8]{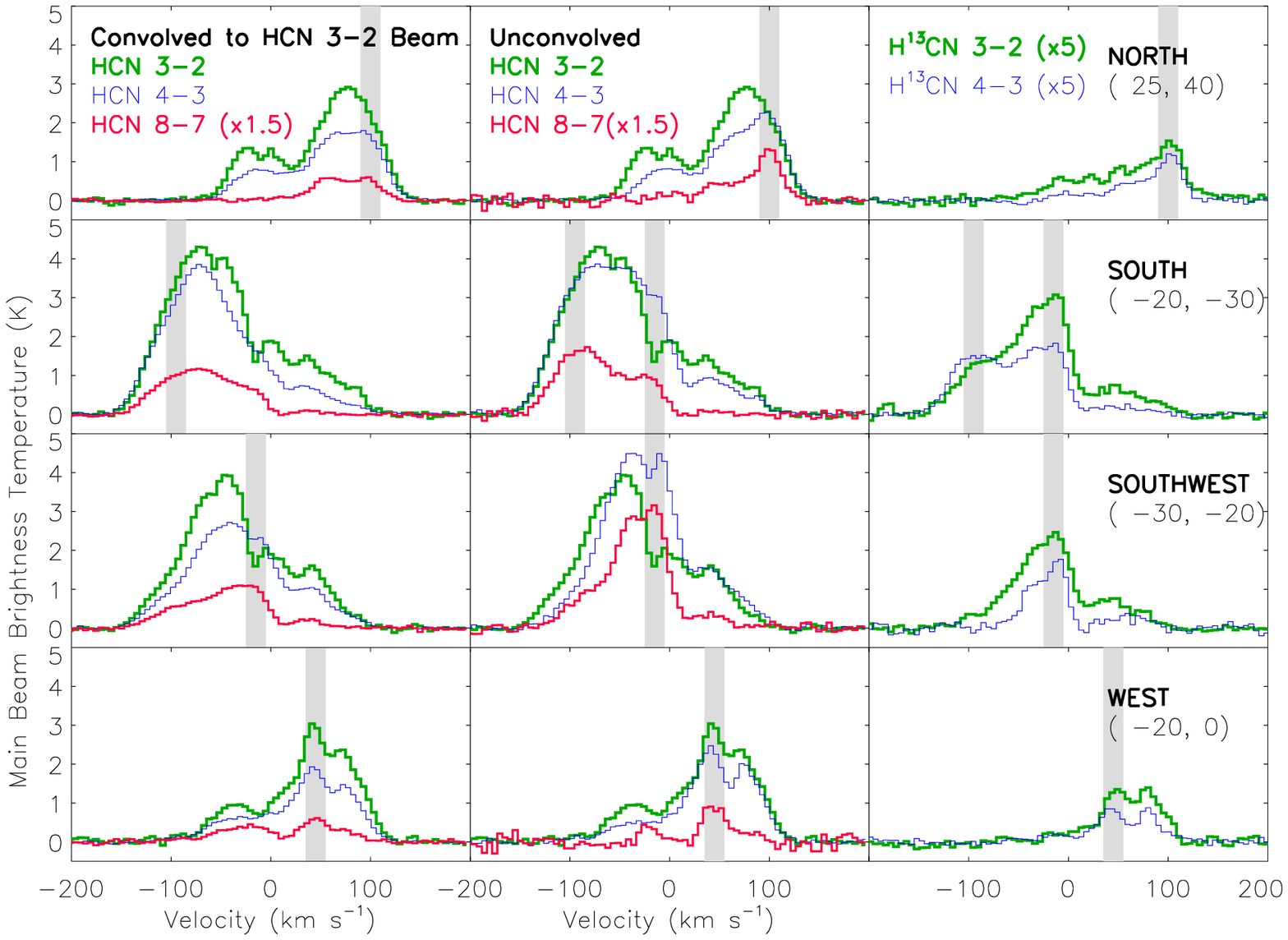}
\vspace{1.3cm}
\caption{ Line profiles of HCN (Left, Middle) and H$^{13}$CN (Right) toward four positions in the CND, from the top: North, South, Southwest, and West.  Spectra have been extracted from maps convolved to the beam size of the 3--2 observations (23.6''). The grey-shaded regions represent the narrow velocity ranges corresponding just to the peaks of individual features listed in Table \ref{Clumps}.}
\label{HCN}
\end{figure*}

\begin{figure*}[tbh]
\hspace{2.2cm}
\includegraphics[scale=0.8]{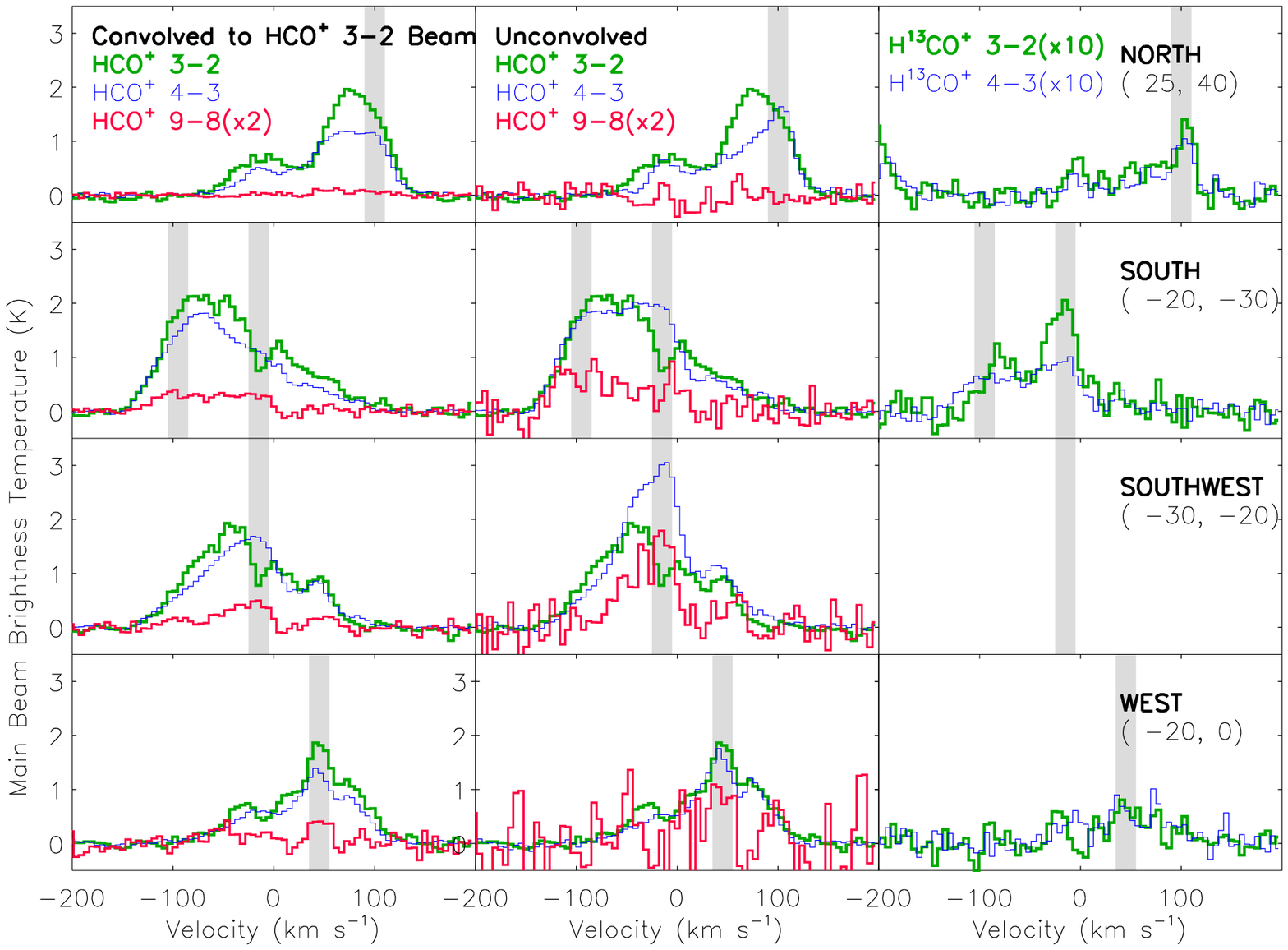}
\vspace{1.3cm}
\caption{ Line profiles of \hco\, (Left, Middle) and H$^{13}$CO$^+$ (Right) toward four positions in CND, from the top: North, South, Southwest, and West.  Spectra have been extracted from maps convolved to the beam size of the 3--2 observations.}
\label{HCO}
\end{figure*}

		Feature\ SW (v = -25 to -5 \kms\hspace{-0.1cm}, offset = -30\arcsec, -20\arcsec) is the source of the strongest HCN 8--7 emission. It also is the location of the strongest H$^{13}$CN 4--3 emission. Emission at the velocities of this feature is seen in both the Southwest and South pointings; we refer to emission at these velocities detected in the South pointing as feature S2 (v = -25 to -5 \kms\hspace{-0.1cm}, offset = -20\arcsec, -30\arcsec), though it is possible that S2 and SW are parts of the same, extended source. The S1 feature (v= -90 to -120 \kms\hspace{-0.1cm}, offset = -20\arcsec, -30\arcsec) is slightly fainter than SW or S2  in HCN 8--7, but is the location of the peak HCN 4--3 emission. Feature N (v = 90 to 115 \kms\hspace{-0.1cm}, offset = +25\arcsec, +40\arcsec) is the faintest of these four (or five) features. Feature W (v = 35 to 55 \kms\hspace{-0.1cm}, offset = -20\arcsec, +0\arcsec) is the location of very strong HCN 3--2 emission.  The velocity range for each of the features is shown shaded in grey in Figures \ref{HCN} and \ref{HCO}.
		 		 
		Although we observed at the frequencies of both the HC$^{15}$N 3--2 and 4--3 lines toward the Southern pointing (corresponding to features S1 and S2), we do not detect either line. In both cases, the observations are somewhat confused by overlap with a nearby strong line.  However, we report upper limits for the peak brightness temperatures of these lines of 0.05 K for HC$^{15}$N 3--2 and 0.03 K for HC$^{15}$N 4--3. 		
				 
	\subsection{Comparison to other studies}
	Interferometric maps of the CND with resolutions of a few arcseconds in lines of HCN 1--0 and 4--3, (\citealt{Chris05} and \citealt{MCHH09}, respectively), resolve the CND into numerous clumps.  Our spectra (Figures \ref{HCN} and \ref{HCO})  allow us to associate the features which we can resolve in velocity space using our single-dish data with spatially-resolved, interferometrically-detected counterparts.
The isotopologue spectra are most useful for this purpose, as they trace the highest column-density gas. This gas generally corresponds well to that detected by the interferometric data, which are less sensitive to faint and extended structure. 

Our feature N corresponds to clump A from \citeauthor{MCHH09} (see their Figure 4). Gas from this clump peaks at a central velocity of +100 \kms\hspace{-0.1cm}. Our feature S corresponds to their clump Q, which peaks at a central velocity of -110 \kms\, and is one of the brightest sources of HCN 4--3 emission detected by \citeauthor{MCHH09} The observed profiles of HCN 8--7 and the isotopologues (Figure \ref{HCN}) toward our  feature W are most similar to the spectrum of clump K from \citeauthor{MCHH09} Clump H of \citeauthor{MCHH09} also lies on the edge of the West pointing and can be seen in the HCN 8--7 maps (Figures \ref{contour} and \ref{87_contour}), however the line profile of clump H in \citeauthor{MCHH09} (which is very narrow and peaks at 60 \kms\hspace{-0.1cm}), does not match our observed isotopologue profiles. This is likely because clump H lies at the very edge of the beam of the West isotopologue pointing, and emission from this source is not well sampled.  
 
Our feature SW corresponds to clump N from \citeauthor{MCHH09} Their line profile for clump N is double-peaked, with a dip at a velocity of  -20 \kms\, that \citeauthor{MCHH09} ascribe to missing short baselines. We see the same dip in our HCN and \hco\, 3--2 and 4--3 line profiles (Figures \ref{HCN} and \ref{HCO}, third row), and significantly, the line profiles of H$^{13}$CN and \hcoiso\, peak at the central velocity of the dip, indicating the main lines suffer from self-absorption. As the line profile of HCN 8--7 peaks at the same velocity, this self-absorption is likely related to the dense CND gas, and not to foreground gas along the line of sight. Absorption at this velocity can also be seen in the CS 7--6 spectrum of \citeauthor{MCHH09} for this clump, as well as in spectra of CO 6--5 and 7--6 from RT12. Absorption at the same velocity is also seen in our spectra of HCN and \hco\, in the South pointing, toward feature S2. The position of clump N from \citeauthor{MCHH09} puts it $\sim$ 14\arcsec\, from the center of the South pointing, placing this clump at the very edge of the beam of our South pointing, consistent with feature S2 and feature SW being the same source, with their emission sampled at slightly different positions.

	\subsection{Line intensities}
 Before measuring the line intensities, we correct the maps for the varying beam sizes by spatially convolving all maps to the resolution of the HCN J = 3--2 observations and interpolating so that the pixelization of the maps is identical. All of the data are also convolved in velocity to a common resolution of 5 \kms\hspace{-0.1cm}.  
We report line intensities which are integrated over the majority of the line profile, avoiding emission from other contaminating features such as the nearby `50 \kms\hspace{-0.1cm}' cloud (M-0.02-0.07) and the southern streamer,  an extension of the `20 \kms\hspace{-0.1cm}' cloud (M-0.13-0.08) identified by \cite{Coil99} that may be interacting with the CND. The chosen velocity ranges for each line profile and the resulting integrated intensities are also given in Table \ref{Intensity}. 

 For each of the four pointings, the line intensities are also integrated over velocity ranges (width $\sim$ 20 \kms\hspace{-0.1cm}) corresponding to interferometrically-detected clumps from \cite{Chris05} and \cite{MCHH09}, given in Table \ref{Clumps}. The velocity range of each feature is chosen to correspond to velocities where the clump is isolated and there is minimal confusion from nearby clumps at similar velocities. The resulting integrated intensities toward all of the features we identify are also reported in Table \ref{Intensity}.

 Figures \ref{HCN} and \ref{HCO} show the spectra extracted from the spatially-convolved maps (lefthand column) and unconvolved maps (central column) at each position in the CND. The velocity range corresponding to individual features is shaded in grey. The noise in each spectrum is much higher for the West position than for other positions in the HCN 8--7 and \hco\, 9--8 lines, as this pointing lies near the edge of the map. All of the \hco\, 9--8 spectra also suffer from increased noise. 
   
 	\subsection{The detection of the vibrationally-excited $v_2=1f$, $J$= 4--3 transition of HCN}
	\label{vibdetect}
\begin{figure*}[t]
\includegraphics[scale=0.4]{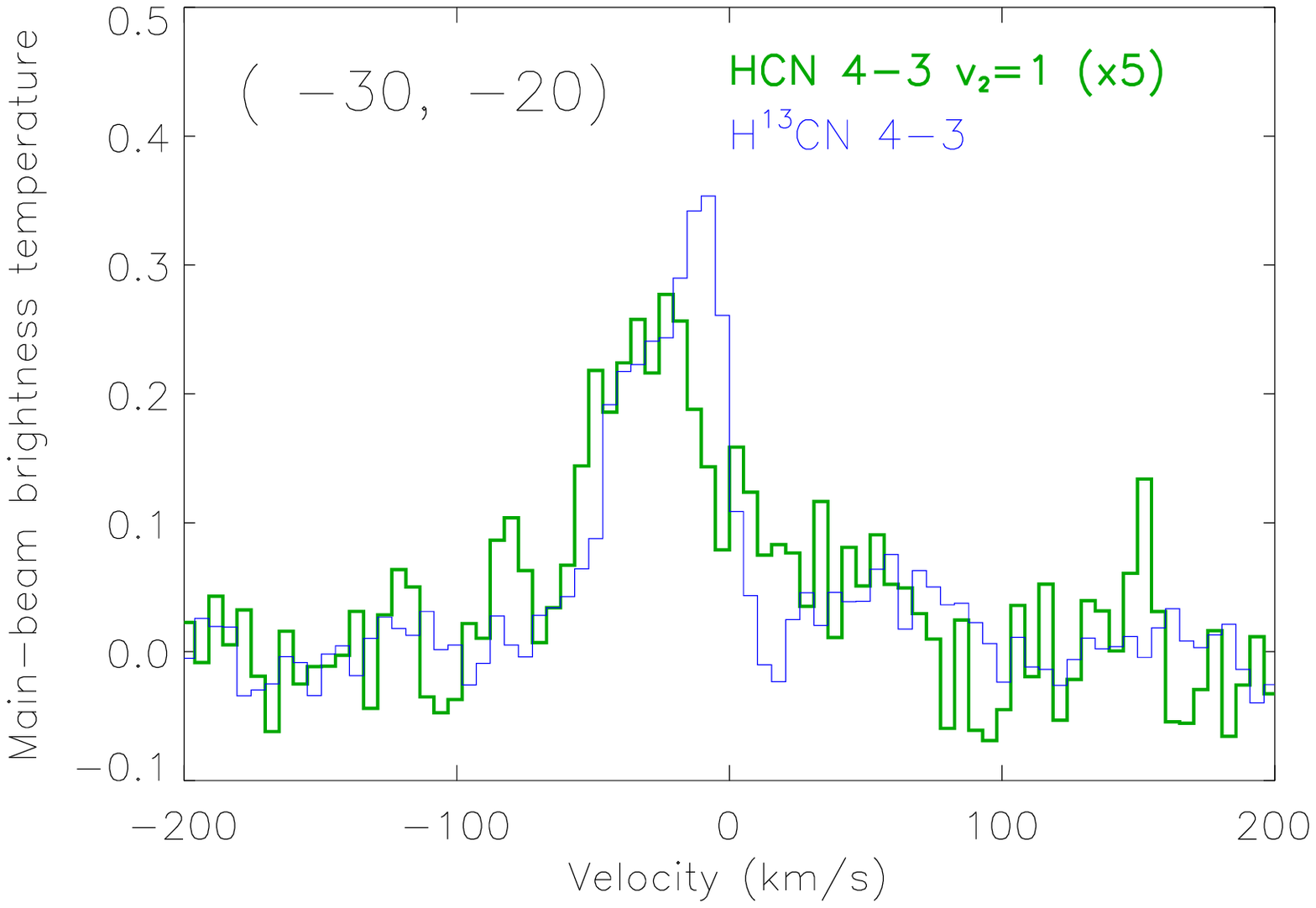}
\caption{ The $J$ = 4--3, $v_2=1$ line of HCN detected toward the Southwest emission peak of the CND. The spectrum of the H$^{13}$CN $J$ = 4--3, $v_2=0$ line toward the same position is superposed for comparison.}
\label{vib}
\end{figure*}

In addition to the transitions already mentioned, we also detect the $J$ = 4--3, $v_2$=1$f$ line of HCN ($\nu$= 356.256 GHz) toward the southern emission peak of the CND (in our Southwest pointing). This is the first detection of vibrationally-excited HCN in the CND. The $v_2$=1 transition corresponds to the bending mode of HCN, and is the lowest-energy of the vibrational modes of this molecule. This line is a doublet, however the $v_2=1e$ line at 354.460 GHz is strongly blended with emission from the main HCN 4--3 line and is not detected. The upper-level energy of the $v_2=1f$ $J$ = 4--3 transition of HCN is 1067.1 K.
The line spectrum is shown in Figure \ref{vib}. The peak intensity of the line is $\sim$50 mK, and it has an integrated brightness temperature of 2.28$\pm$0.23 \kms\hspace{-0.1cm}. The peak of this line lies at a velocity of -20 \kms\hspace{-0.1cm}, associating it with the S2/SW clump.

	\subsubsection{Properties of the vibrationally-excited emission}
	By comparing the upper-level column densities of the HCN $J$=4--3 $v_2$=0 and $v_2$=1 lines, we can derive the vibrational excitation temperature for feature S2/SW (technically, just for the SW component of this clump, as we do not have a good observation of the vibrationally-excited line from the South pointing).  
	
	We assume that the HCN $J$=4--3 $v_2$=1 line is optically thin. As the HCN $J$=4--3 $v_2$=0 line is not optically thin, we determine the integrated intensity of this line by scaling the H$^{13}$CN 4--3 line intensity (assumed to be optically thin) by the $^{12}$C/$^{13}$C ratio of 25. The column densities are then determined using:
	
	\begin{equation} N_{\textrm{\small HCN}} = \frac{8\pi k\nu^2}{A_{ul}hc^3}\int \textrm{T}_{\textrm{MB}} dv,
	\label{thick}
	\end{equation}

	where $k$ is Boltsmann's constant ($1.380658\times10^{-16} erg K^{-1}$), $h$ is Planck's constant ($6.626076\times10^{-27}$ erg s), A$_{(4,3,0)} = 2.054\times10^{-3}$ s$^{-1}$ is the Einstein A-coefficient for the J=4--3, $v_2=0$ transition \citep{Dumouchel10}, taken from the Leiden Atomic and Molecular Database \citep[LAMDA,][]{Schoier05}, A$_{(4,3,1)} =1.876\times10^{-3}$  s$^{-1}$ for the J=4--3, $v_2=1$ transition \citep{Harris}, $\nu_0$ is the frequency at line center, and $\int$T$_{\textrm{MB}} dv$ is the velocity-integrated main beam brightness temperature. 
	
Solving for the vibrational excitation temperature, we find T$_{\mathrm{ex}}$ = T$_{vib} = 205 \pm$ 10 K for feature S2/SW. If, however, the H$^{13}$CN 4--3 line is not optically thin, the value we derive would be an overestimate of the true T$_{vib}$. 

	\subsubsection{Collisional vs. Radiative excitation of the $v_2=1$ line}
We consider two possibilities for the excitation of this line: either collisional excitation, due to extremely high volume densities, or radiative excitation through the rovibrational transitions of HCN at 14 $\mu$m. For collisional excitation to excite this transition requires the density to be at least comparable to the critical density of the $v_2$=1 transition, which is $\sim5\times10^{11}$ \cm\, \citep{Ziurys}.  As we will show in Section \ref{Res}, such a density is more than four orders of magnitude above the highest densities that we constrain in the CND, making purely collisional excitation an unlikely source of excitation. More likely is that HCN is radiatively excited, either externally (via a sufficiently strong 14 $\mu$m background radiation field), or internally, via sufficiently hot dust mixed with the gas.
		 
		 Radiative excitation of HCN will not just populate the $v_2=1$ states, but will also affect the populations of the rotationally-excited levels in the ground vibrational state. In the presence of a strong 14 $\mu$m radiation field, molecules in a given rotational state have a statistical likelihood to be vibrationally excited, and subsequently decay into a higher rotational level of the ground vibrational state \citep{Morris75,Carroll81}. This pumping shifts the populations of the rotationally excited levels to the higher J levels, mimicking the effect of a higher gas density. This phenomenon was first observed for HCN in the ISM by \cite{Ziurys}, and more recently was suggested to be an important source of excitation for HCN in NGC 4418 \citep{Sakamoto}, based on the detection of the J=4--3, $v_2=1$ line, as well as absorption in the 14 $\mu$m band. The J=4--3, $v_2=1$ line of HCN has also been detected in the z=0.043 AGN-hosting galaxy IRAS 20551Ð4250 \citep{Imanishi}, however it is unclear to what extent radiative pumping may be affecting the populations of the $v_2=0$ states in this source. Radiative pumping has also been suggested to be important for other molecules, such as CS in hot protostellar cores \citep{Hauschildt93, Hauschildt}, HNC in luminous infrared galaxies \citep{Aalt07a}, and, most recently, HF, HCl, SiO and CO, in the presence of a sufficiently strong near-IR, optical or UV radiation field \citep{GC13}.
		  
Given the vibrational temperature we infer from our HCN observations, we can determine which levels of the ground vibrational state would be affected by radiative pumping. The criterion for effective radiative pumping is given by Equation (2) of \cite{Sakamoto}, reproduced here:
	
	\begin{equation} e^{-T_0/T_{\mathrm{vib}}} \mathrm{A}_{\mathrm{vib}} \geq \mathrm{A}_{\mathrm{rot} J},
	\label{pump}
	\end{equation}	
	
where T$_0$ is the level energy of the $J$=4, $v_2=1$ state, T$_{\mathrm{vib}}$ is our computed vibrational excitation temperature, A$_{\mathrm{vib}}$ is the Einstein A for the rovibrational transition which links the $J$=4 $v_2=0$ and $v_2=1$ states, equal to $\sim$2.3 s$^{-1}$ \citep{Harris}, and A$_{\mathrm{rot} J}$ is the Einstein-A coefficient for $J$= 4--3, $v_2$=0. 
	
Given our observed T$_{vib}$, Equation \ref{pump} shows that pumping of the $v_2=0$ rotational transitions of HCN via the rovibrational transitions should be efficient up to $J\sim$12 for feature S2/SW, indicating the level populations of all observed lines in this feature (both $v_2=0$ and 1) are likely affected by radiative pumping.

\section{Excitation Analysis}
\label{Ana}
	For the remainder of this paper, we focus on excitation analyses of the lines of HCN and \hco\, in the CND, both for the majority of the line profile as well as for selected velocity intervals corresponding to previously identified clumps in the CND, listed in Table \ref{Clumps}.  Again, we separately report both emission from feature S2 observed at ($-20'',-30''$) and feature SW observed at ($-30'',-20''$), although these are likely part of the same structure.  The resulting velocity-integrated main beam brightness temperatures for five transitions of both HCN and \hco\, for each feature (Table \ref{Intensity}) are sufficient to constrain a fit to single gas component characterized by four parameters: the temperature, H$_2$ density, molecular column density, and beam filling factor, with a fifth parameter, [$^{12}$C]/[$^{13}$C], set to a fixed value of 25 \citep[][]{WR94,Wilson99,Riq10}. 

		To fully model the line radiative transfer of HCN and \hco, we also consider contributions to the radiation field due to the local and global radiation background.  In addition to the cosmic microwave background, emission from warm dust in the CND gives rise to a continuum background at the frequencies of the rotational lines of HCN and \hco. Both HCN and \hco\, also have rovibrational transitions in the near and mid-infrared (the lowest energy modes for both are the bending modes, occurring at 14 and 12 $\mu$m, respectively). Our initial analysis reveals that radiative excitation through these rovibrational transitions is the most likely source of excitation of the detected $v_2=1f$ $J$= 4-3 line of HCN (see Section \ref{vibdetect}), making it important to include an accurate description of the radiation field in the CND. 
		
	We first use the statistical equilibrium radiative transfer code RADEX \citep{vanderTak07}, a zero-dimensional non-LTE code, which employs the escape probability formalism to model the observed line intensities as a function of the physical conditions in the source. The escape probability method simplifies the radiative transfer calculation by assuming that photons either completely escape the source (the likelihood of this is dependent on the local opacity, which is itself determined by the source geometry), or are immediately absorbed at the same location where they were emitted. A further simplification employed in the above method is the assumption of uniform physical conditions throughout the source.
	
	In addition, in Section \ref{rat} we also compare the RADEX results for HCN to those from a more sophisticated radiative transfer code \citep[RATRAN:][]{Hog00} which takes into account the internal radiation field due to embedded dust. 
		
	\subsection{RADEX } 
	\subsubsection{Input Parameters}
	The radiative and collisional coefficients were obtained from LAMDA for the rotational lines of both HCN and \hco\, \citep{Dumouchel10,Flower99}. We also used the radiative excitation rates of the vibrationally-excited lines of HCN  from \cite{Thorwirth03} to model the observed $v_2$=1 line. As the collisional coefficients of the vibrationally-excited transitions are unknown, we do not take into account any collisional excitation of the vibrationally-excited states, consistent with our previous conclusion that collisions do not contribute to the excitation of this line. We assume in our analysis of the collisional excitation of the rotationally-excited levels that H$_2$ is the main collisional partner for HCN and \hco.  
	
	For fits to the intensities of individual features, we assume a velocity full-width half maximum (FWHM) of 20 \kms\,\hspace{-0.1cm} for the determination of the escape probability, consistent with values of $15-50$ \kms\, measured by interferometric studies for the corresponding clumps \citep{Chris05, MCHH09}. For fits to the  majority of the profile, we use a larger FWHM of 50-100 \kms\hspace{-0.1cm}. The assumed geometry for determining the escape probability is that of an expanding sphere (equivalent to the large velocity gradient or LVG approximation). Changing the assumed geometry to that of a uniform sphere, slab, or turbulent medium does not significantly alter the results presented here. 
	
	For the radiation field at the wavelength of the rovibrational transitions of HCN and \hco, we adopt  the mid-infrared spectrum as measured by ISO. Only the observations centered on the nucleus, Sgr A*, have been published \citep{Lutz}, however an additional spectrum of the southern emission peak of the CND is available from the NASA-IRAC Infrared Science Archive\footnotemark[1]\footnotetext[1]{http://irsa.ipac.caltech.edu/} (IRSA) toward an offset of (-1.3$''$,-37.6$''$) from the position of Sgr A*, which we use for this analysis. The calibrated spectrum is not corrected for the variable extinction in the region, and additionally represents the total flux of radiation from a fairly large aperture (14\arcsec\, by 20\arcsec). The local intensity of the ambient radiation field might then differ by a factor of 2 or more from the measured value. 

		\subsubsection{Grids} 
	We use RADEX to construct grids of predicted line intensities over the given range of input temperature, density, and column density. For each set of temperature, density, and column density values, RADEX calculates main line intensities and opacities, from which the isotopologue intensities can also be derived using our assumed [$^{12}$C]/[$^{13}$C] ratio. We fit separately for each molecule (HCN and \hco). 
	The line intensities generated by RADEX for all five lines of each species can then be compared to the measured line intensities by introducing an additional parameter: a beam filling factor for the emitting region. Given the measured uncertainties on the line intensities, one can then determine the (reduced) chi-squared parameter for the fit of the measured to the modeled line intensities for the entire range of temperature, density, and column density considered.   
	
	\subsubsection{Fitting Constraints}
	We impose several constraints on our model to eliminate unphysical solutions. First, as ammonia temperature measurements indicate that gas temperatures in the CND are at least as hot as 50 K \citep{McGary01}, we exclude lower temperatures from our input parameter grid. 
	We also find that where column densities are in excess of $10^{16}$ cm$^{-2}$, and densities are between $10^5$ and $10^6$ \cm\, the HCN 1--0 line undergoes strong maser action. HCN 1--0 masering is not observed \citep{Chris05}  and so this region of parameter space is also excluded. 
	
	\subsubsection{Models}
		We fit the line intensities integrated over the majority of the line profile at each position (North, South, Southwest, and West) with a single temperature and density model.  We consider temperatures in the range 50 to 600 K, densities in the range of $10^4$ to $10^8$ \cm, and column densities in the range of $10^{14.5}$ to $10^{16.5}$ cm$^{-2}$ for HCN and $10^{13}$ to $10^{15}$ cm$^{-2}$ for \hco. In addition, we perform fits over the same range of physical parameters to line intensities from each of the identified features (N, S2, S2/SW, and W) integrated over the velocity ranges given in Table \ref{Intensity}. 
		
\section{Results of the RADEX Excitation Analyses}
\label{Res}	
	\subsection{Fits to the majority of the line profile}
\begin{figure*}[tbh]
\hspace{-0.5cm}
\includegraphics[scale=0.4,angle=270]{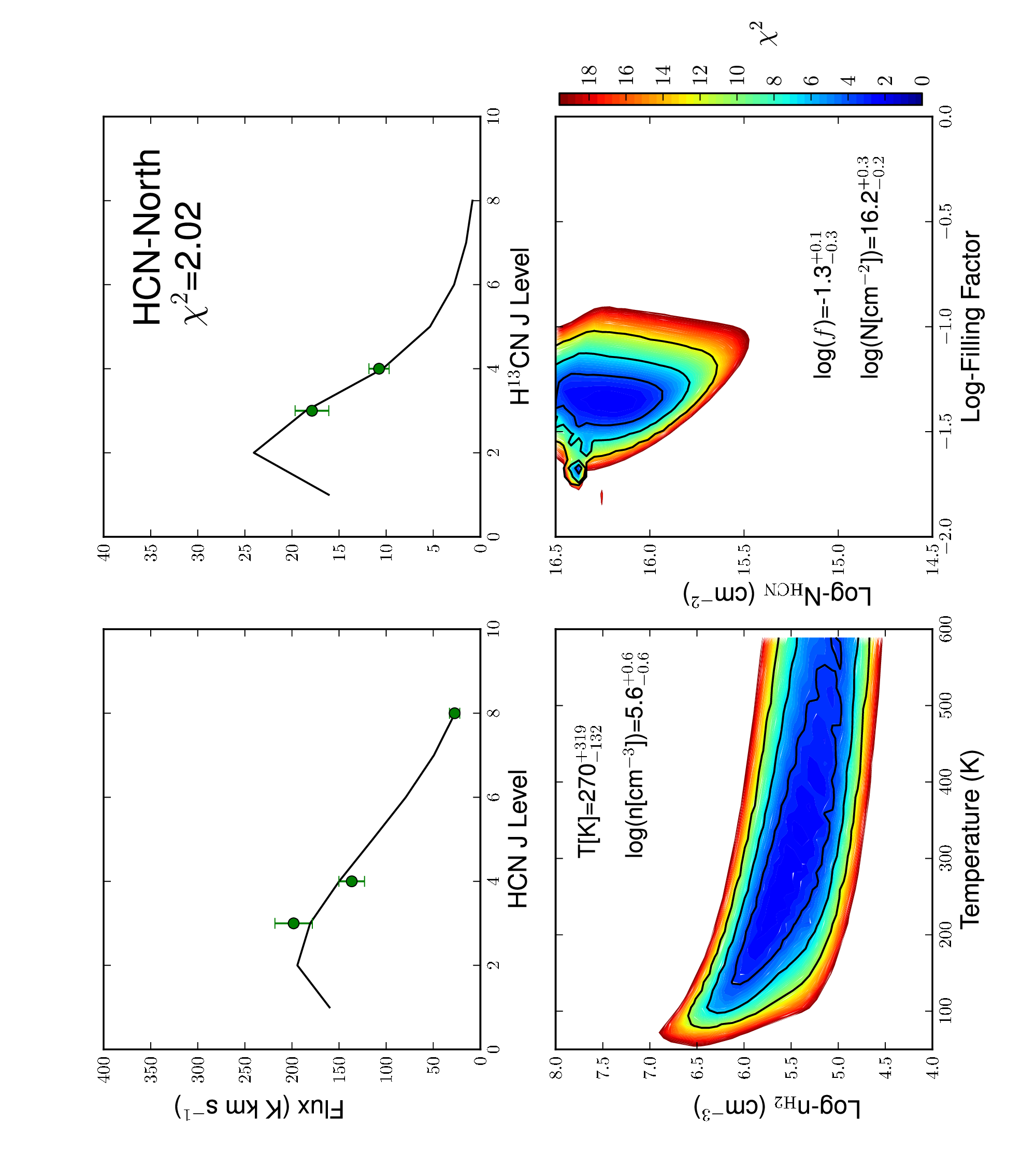}
\includegraphics[scale=0.4, angle=270]{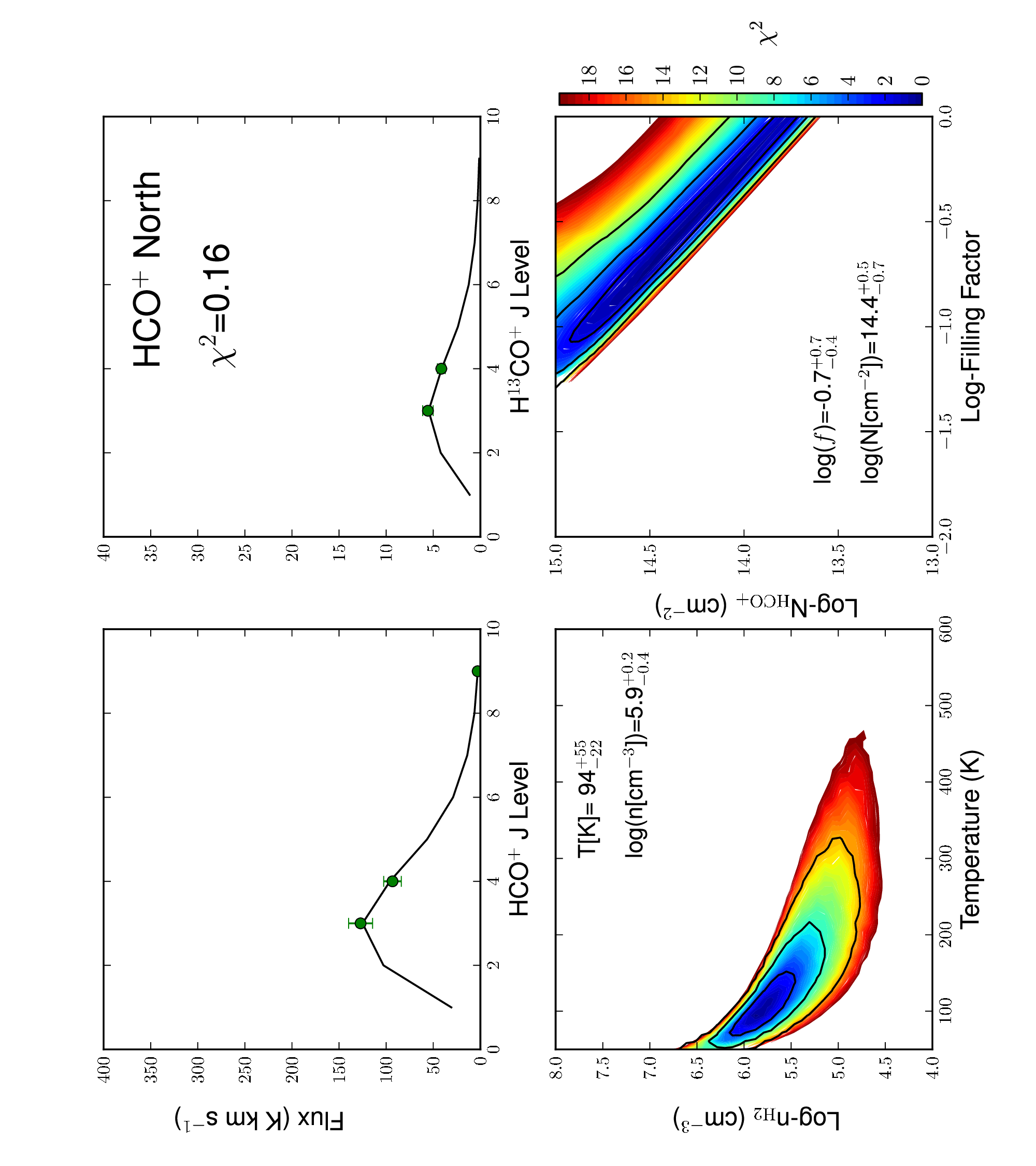}
\caption{ Chi-squared fits to a 1-component model of HCN and \hco\, excitation for line intensities toward the North position in the CND (integrated over the majority of the line profile). Top Row: Boltzmann plots for (left to right) HCN, H$^{13}$CN, \hco, and \hcoiso, showing the best fit solution. Bottom row: two dimensional likelihood distributions, with contours of the 1-, 2-, and 3-$\sigma$ deviations from the most likely value over the full grid of temperatures and densities considered as well as filling factor and column density. } 
\label{hcn_north}
\end{figure*}

\begin{figure*}[tbh]
\hspace{-0.5cm}
\includegraphics[scale=0.4, angle=270]{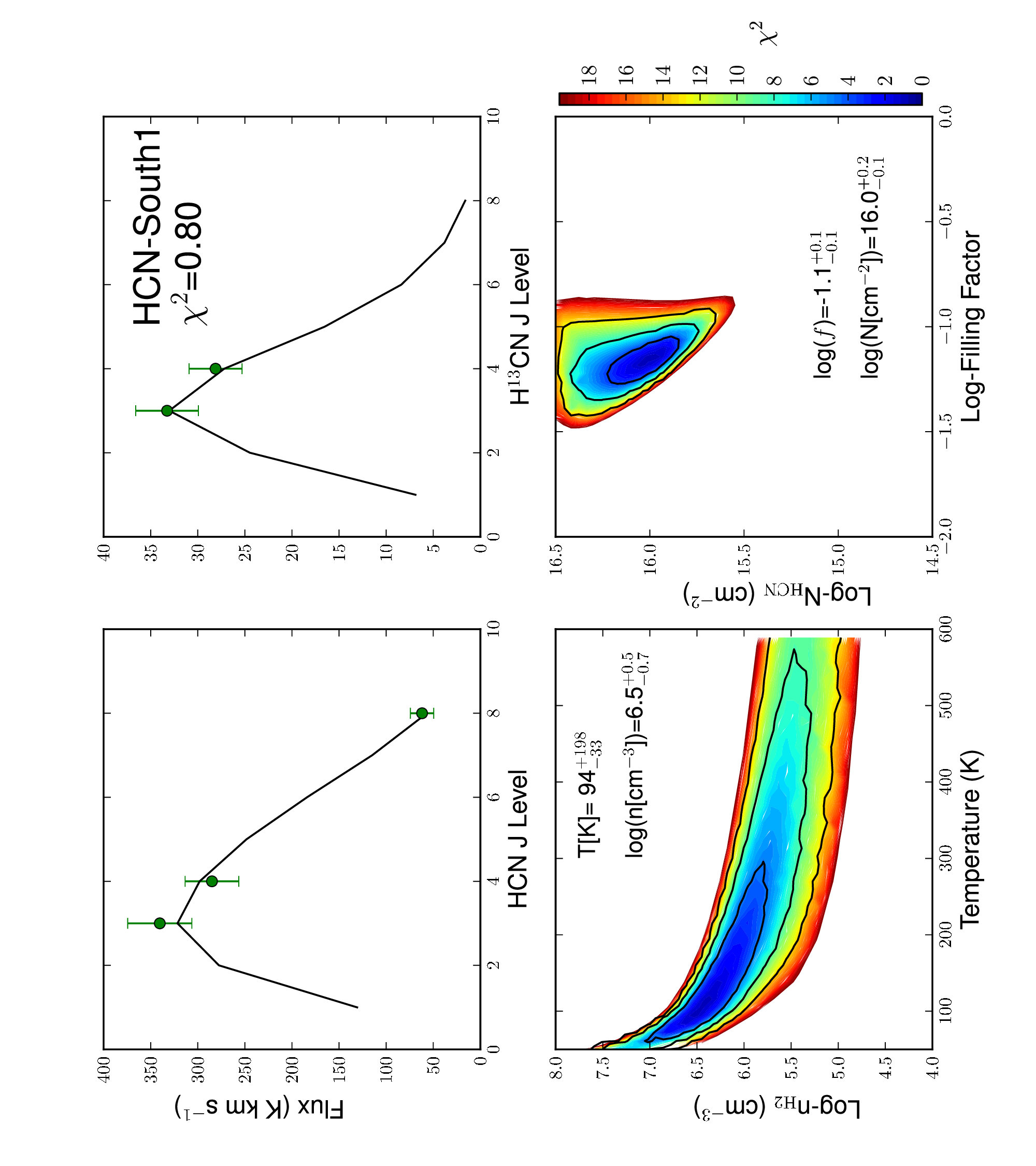}
\includegraphics[scale=0.4, angle=270]{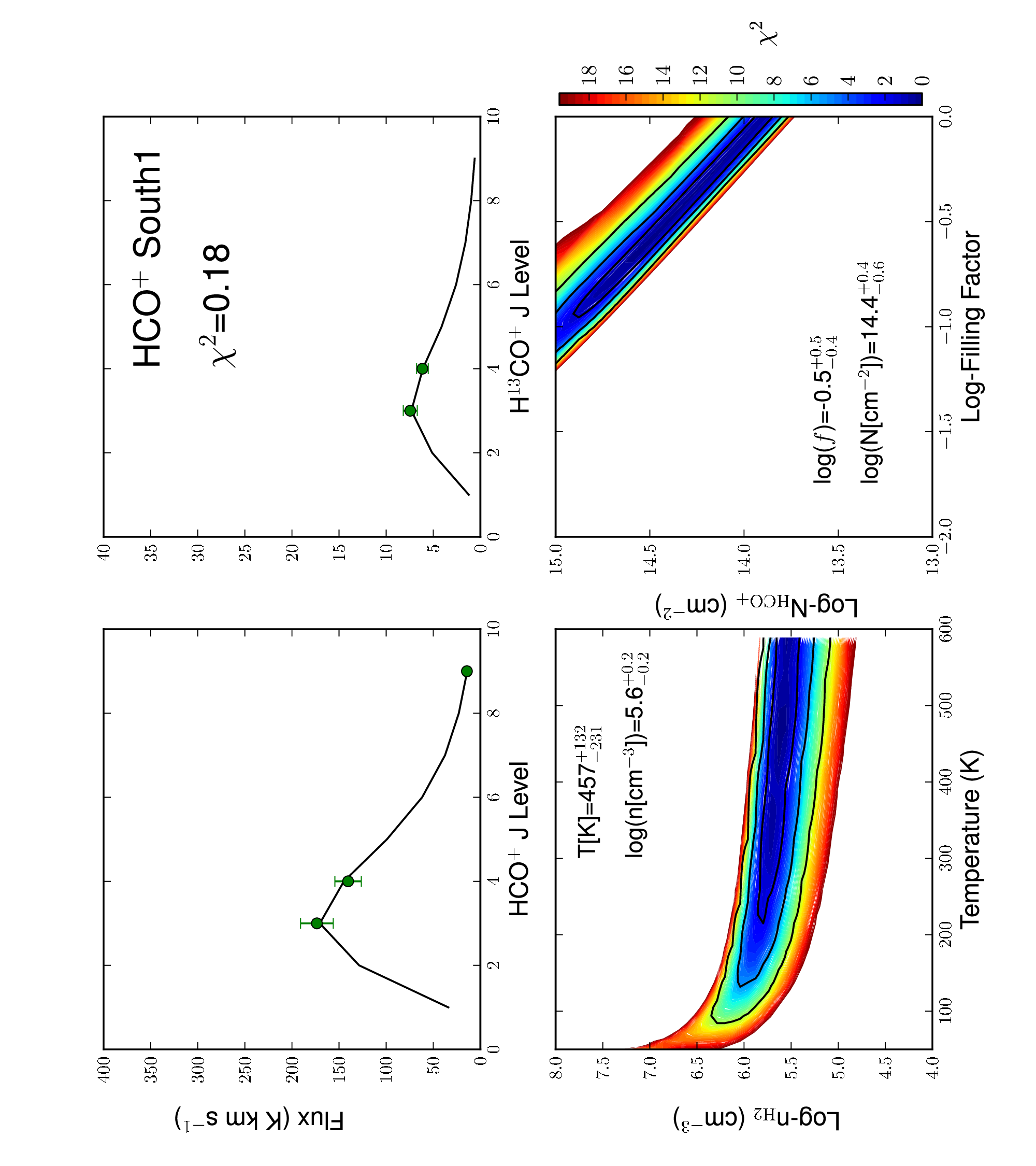}
\caption{Chi-squared fits to a 1-component model of HCN and \hco\, excitation for line intensities toward the South-1 position in the CND (integrated over the majority of the line profile). Rows are the same as for Figure \ref{hcn_north}.}
\label{hcn_south}
\end{figure*}

\begin{figure*}[tbh]
\hspace{-0.5cm}
\includegraphics[scale=0.4, angle=270]{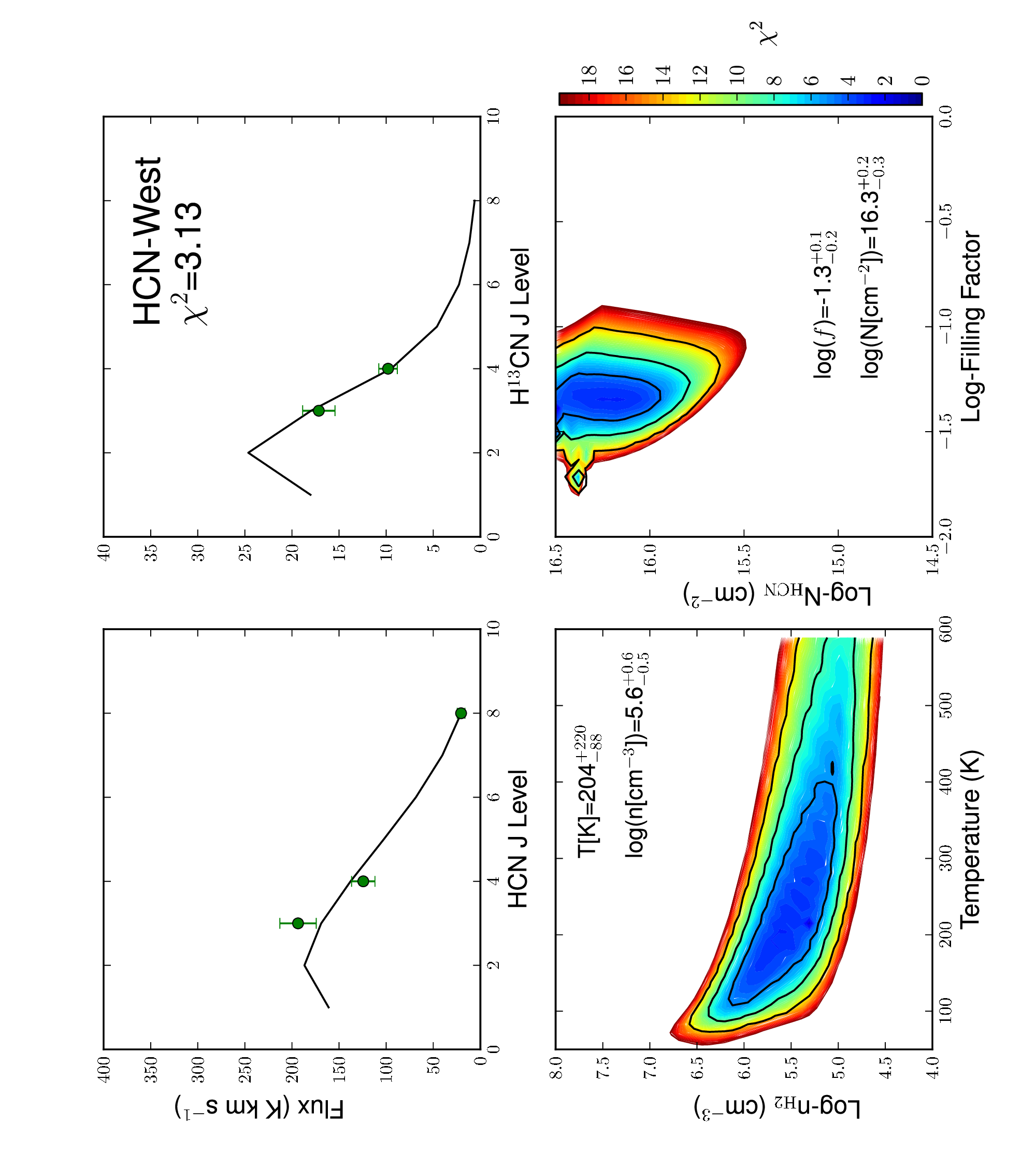}
\\
\\
\caption{Chi-squared fits to a 1-component model of HCN excitation for line intensities toward the West position in the CND (integrated over the majority of the line profile). Rows are the same as for Figure \ref{hcn_north}.}
\label{hcn_west}
\end{figure*}

\begin{figure*}[tbh]
\hspace{-0.5cm}
\includegraphics[scale=0.4, angle=270]{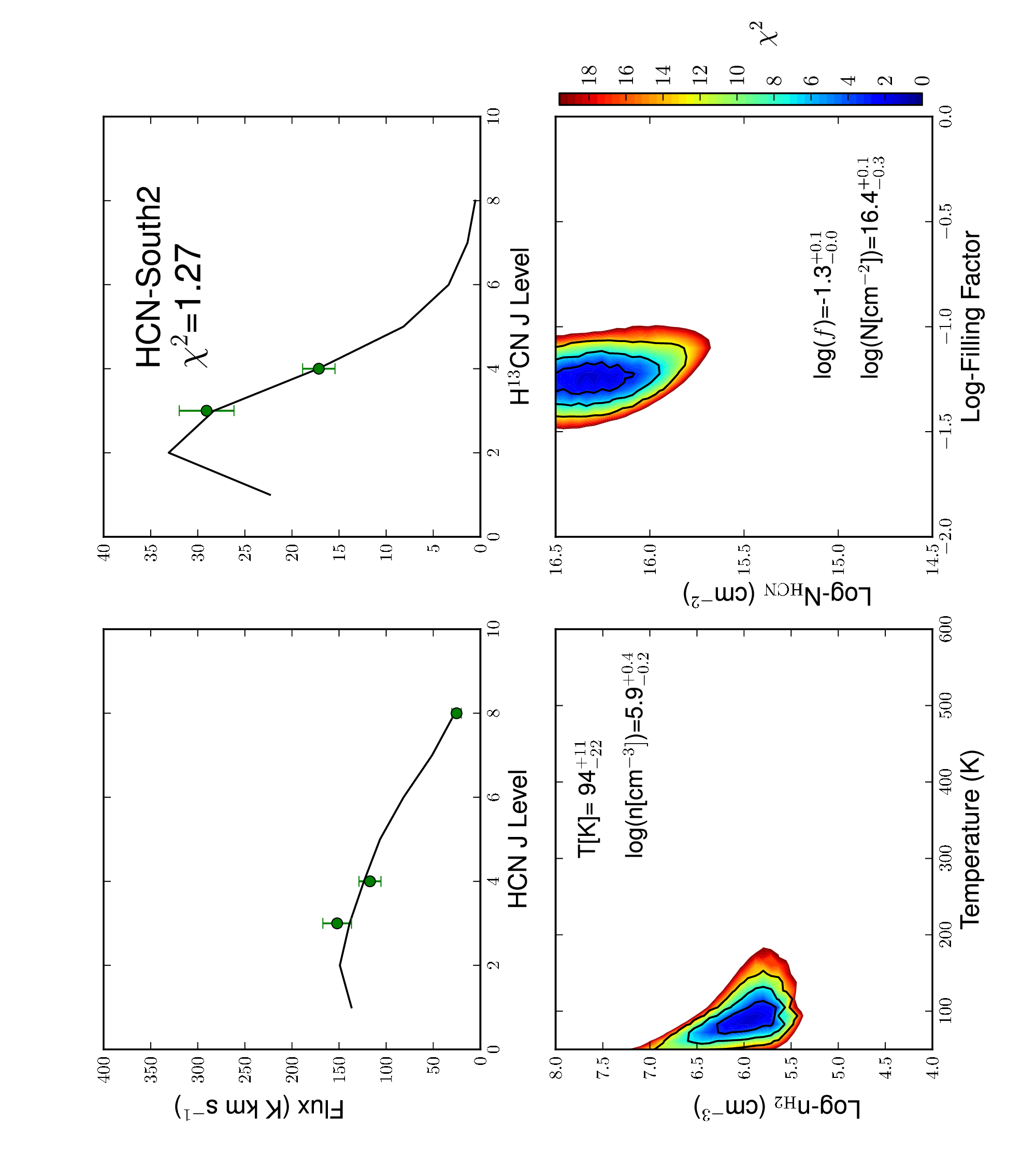}
\includegraphics[scale=0.4, angle=270]{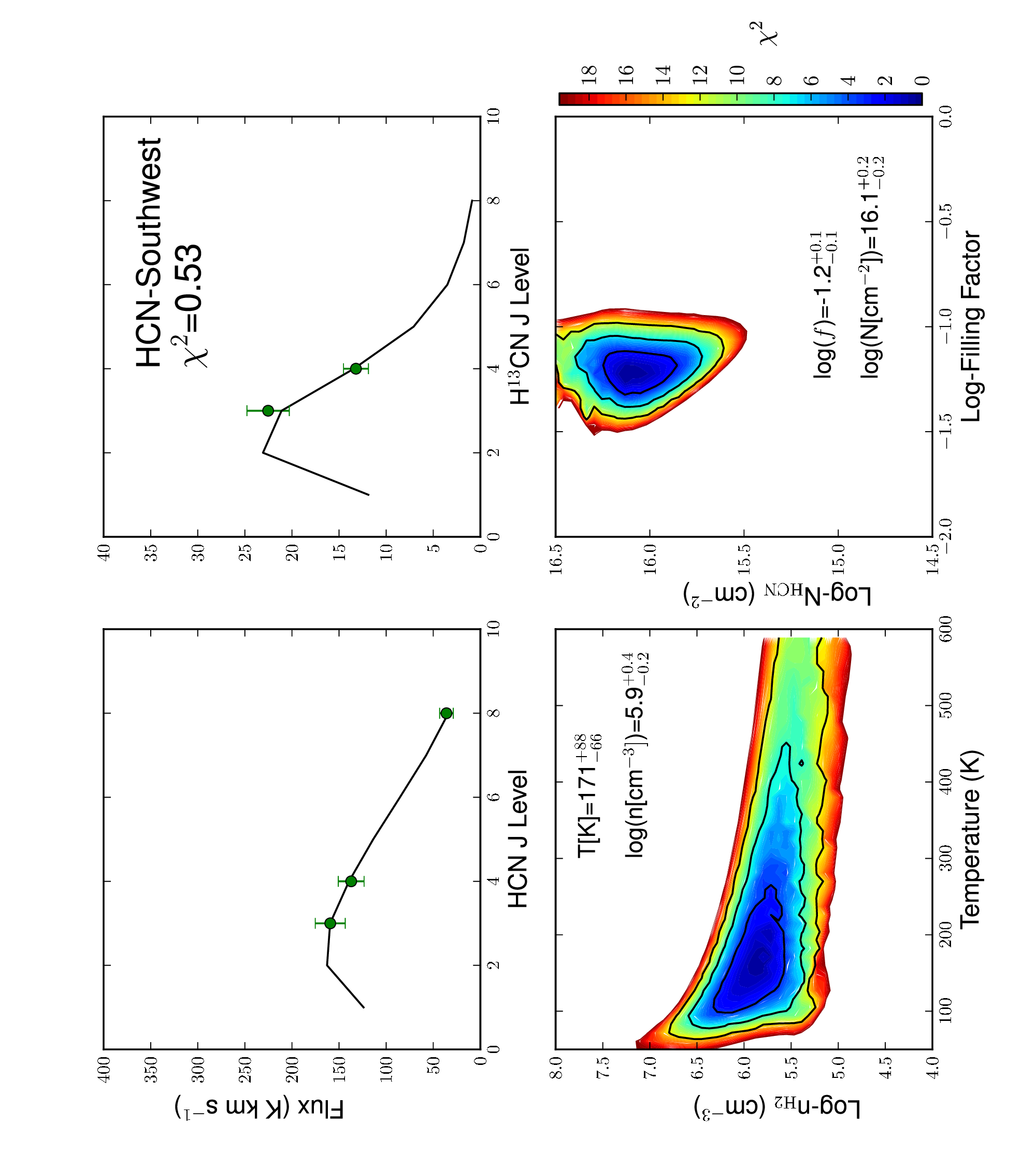}
\\
\\
\caption{Chi-squared fits to a 1-component model of HCN excitation for line intensities toward the South-2 and Southwest positions in the CND (integrated over the majority of the line profile). Rows are the same as for Figure \ref{hcn_north}.}
\label{hcn_southwest}
\end{figure*}

	For the observed CND positions, we find acceptable fits ($\chi^2<5$, where the $\chi^2$ values presented here are equivalent to the reduced chi-squared values, as our fits have a single degree of freedom) to line intensities integrated over the majority of the line profiles for all positions except \hco\, intensities toward the South-2, Southwest and West positions. We show plots of the $\chi^2$ distribution for the fits to HCN and \hco\, emission over the majority of the line profile at all positions in Figures \ref{hcn_north}, \ref{hcn_south}, \ref{hcn_west} and  \ref{hcn_southwest}.  The resulting best-fit density, temperature, column density, and filling factor for each position as well as 1-$\sigma$ errors are reported in Table \ref{Results}. The volume density for typical dense gas in the CND is relatively well constrained by these fits to be between $10^5 - 10^7$ \cm. In contrast, the temperature is more poorly constrained, and varies from 60 K to 600 K, which is the upper limit of the range of temperatures probed by our models. For several cases, including HCN fits to the North and West pointings, and \hco\, fits to the South-1 pointing, the upper bound on the temperature is unconstrained by these fits.

	The volume densities derived from our \hco\, and HCN observations are comparable for the North position, but the densities derived for the South position are not consistent within the uncertainties; the best-fit density derived for this position from HCN  is an order of magnitude higher than that derived using \hco. In contrast, the derived temperatures are consistent for the South position, but the temperature derived from \hco\, is significantly cooler for the North position than that derived using HCN. 
	
	The derived filling factors for both positions also differ significantly: the best-fit filling factors for HCN emission toward all positions are less than 0.08, while they are substantially greater for \hco\, (Table \ref{Results}). This suggests either that the HCN emission in the CND is significantly more clumpy than \hco\, emission, or \citep[as interferometric maps show that the two species exhibit similar small-scale structure;][]{Chris05} that more of the \hco\, emission detected in single-dish observations originates in an extended gas component. If the extended component has different excitation conditions than the gas which is predominantly traced by HCN, this could also explain the fact that we are unable to fit the \hco\, intensities well with a single excitation component for three positions in the CND (South-2, Southwest, and West).
	
	The best-fit HCN column densities are somewhat higher than mean HCN column densities of $\sim10^{15}$ cm$^{-2}$ derived by \cite{Chris05} from observations of HCN 1-0 in individual clumps in the CND. This could either be the result of missing flux in the interferometric observations of \citeauthor{Chris05}, or could indicate that column densities derived from the HCN 1--0 line are underestimated if, for example, the line is more optically thick than assumed. However, our best-fit models predict that the HCN 1--0 line should be generally be optically thin and/or weakly inverted ($\tau<1$), consistent with that recently found by Smith et al. (MNRAS submitted). This is the case for all features except possibly S2 (for which our best fit model predicts $\tau\sim$4), making it more likely that the difference in column densities is due to extended emission which is missed by previous interferometric observations. 
					
\begin{figure*}[tbh]
\hspace{-0.5cm}
\includegraphics[scale=0.5, angle=270]{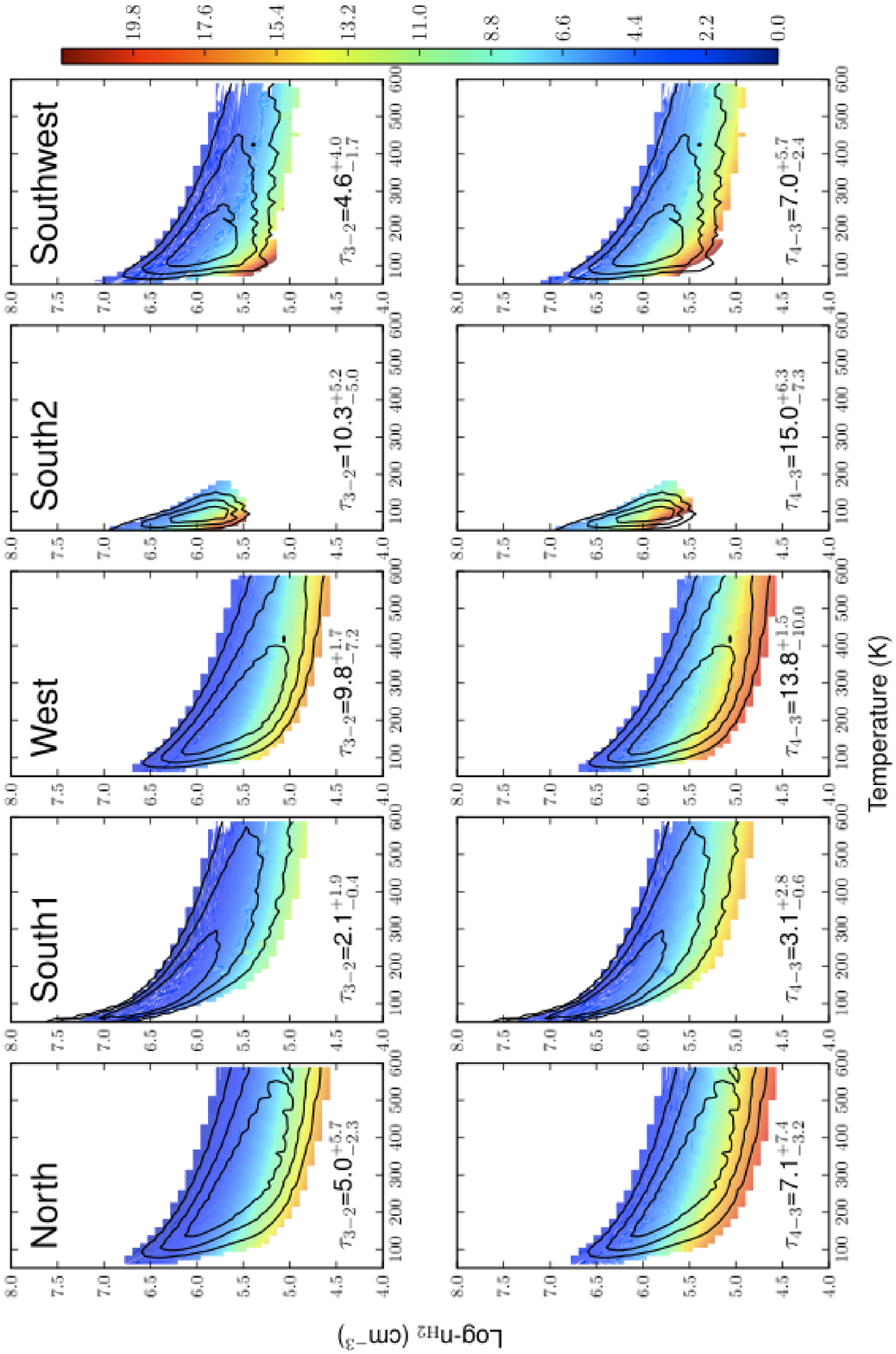}
\\
\\
\caption{HCN line opacities for the 3-2 (Top) and 4-3 (Bottom) transitions, derived from fits to line intensities integrated over the majority of the line profile. Contours show the 1-, 2-, and 3-$\sigma$ deviations from the most likely temperature and density over the full grid of physical conditions that were considered.}
\label{hcn_OP_full}
\end{figure*}

\begin{figure*}[tbh]
\hspace{-0.5cm}
\includegraphics[scale=0.5, angle=270]{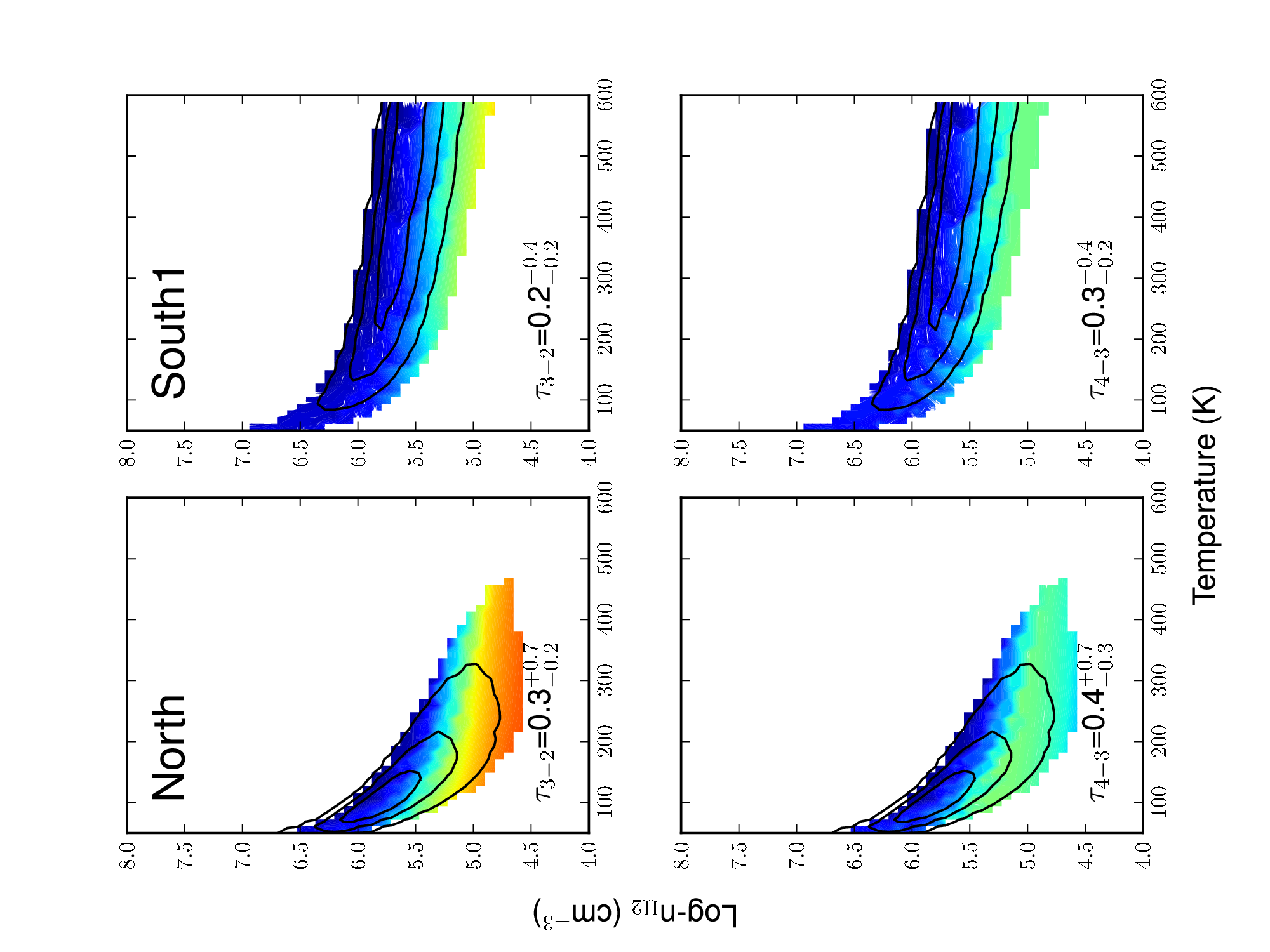}
\includegraphics[scale=0.5, angle=270]{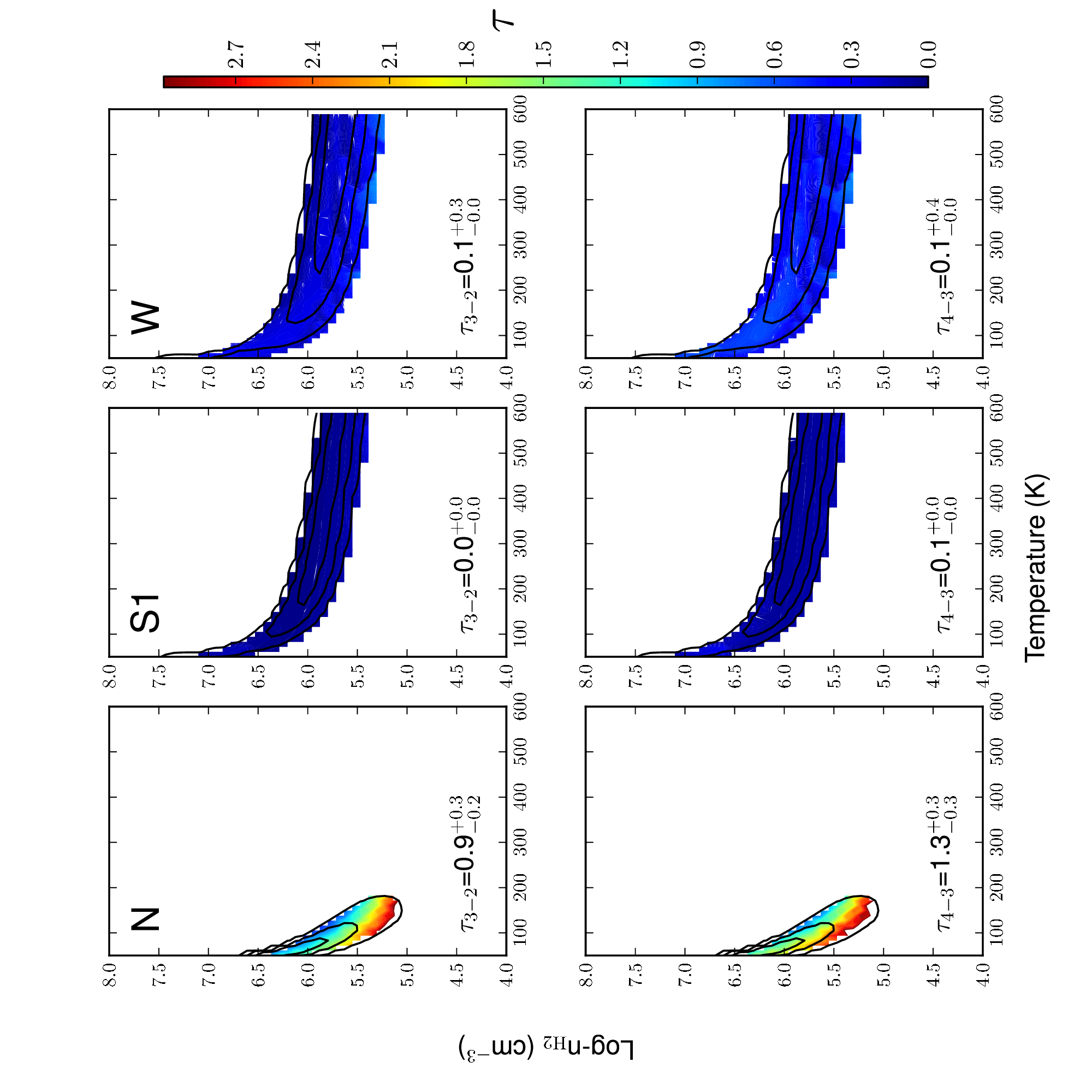}
\\
\\
\caption{\hco\, line opacities for the 3-2 (Top) and 4-3 (Bottom) transitions, derived from fits to line intensities integrated over the majority of the line profile (Left) and fits to individual features (Right). Contours show the 1-, 2-, and 3-$\sigma$ deviations from the most likely temperature and density over the full grid of physical conditions that were considered.}
\label{hco_OP_full}
\end{figure*}

	The  \hco\, 3--2 and 4--3 emission toward the North and South pointings is also found to be optically thin ($\tau < 1$), while the HCN 3--2 and 4--3 emission is everywhere quite optically thick ($\tau > 2$), with HCN 8--7 predicted to be optically thick as well ($\tau > 1$), although we do not observe this transition in H$^{13}$CN.  The distributions of opacities are shown in Figures \ref{hcn_OP_full} and \ref{hco_OP_full}.  Opacities are not determined for \hco\, toward the South-2, Southwest, and West pointings, as acceptable model fits were not found for the line intensities toward those positions. The  \hco\, emission is however likely optically thick toward the South-2 and Southwest positions, as the ratios of H$^{12}$CO$^{+}$\, lines to their $^{13}$C isotopologues are between 10 and 20, significantly less than the assumed [$^{12}$C]/[$^{13}$C] isotope ratio of 25.  It is not possible however to accurately determine the opacity just from the ratio of these line intensities and the intrinsic [$^{12}$C]/[$^{13}$C] isotope ratio, as this also requires knowledge of the excitation temperature for each level. The excitation temperatures for the $^{12}$C and $^{13}$C isotopologues cannot be assumed to be the same; models show that the excitation temperature for the $^{13}$C isotopologue for typical CND conditions can be up to a factor of 2 lower than for the main line. 

		\subsection{Fits to individual features}	
		\label{full}
\begin{figure*}[tbh]
\hspace{-0.5cm}
\includegraphics[scale=0.4,angle=270]{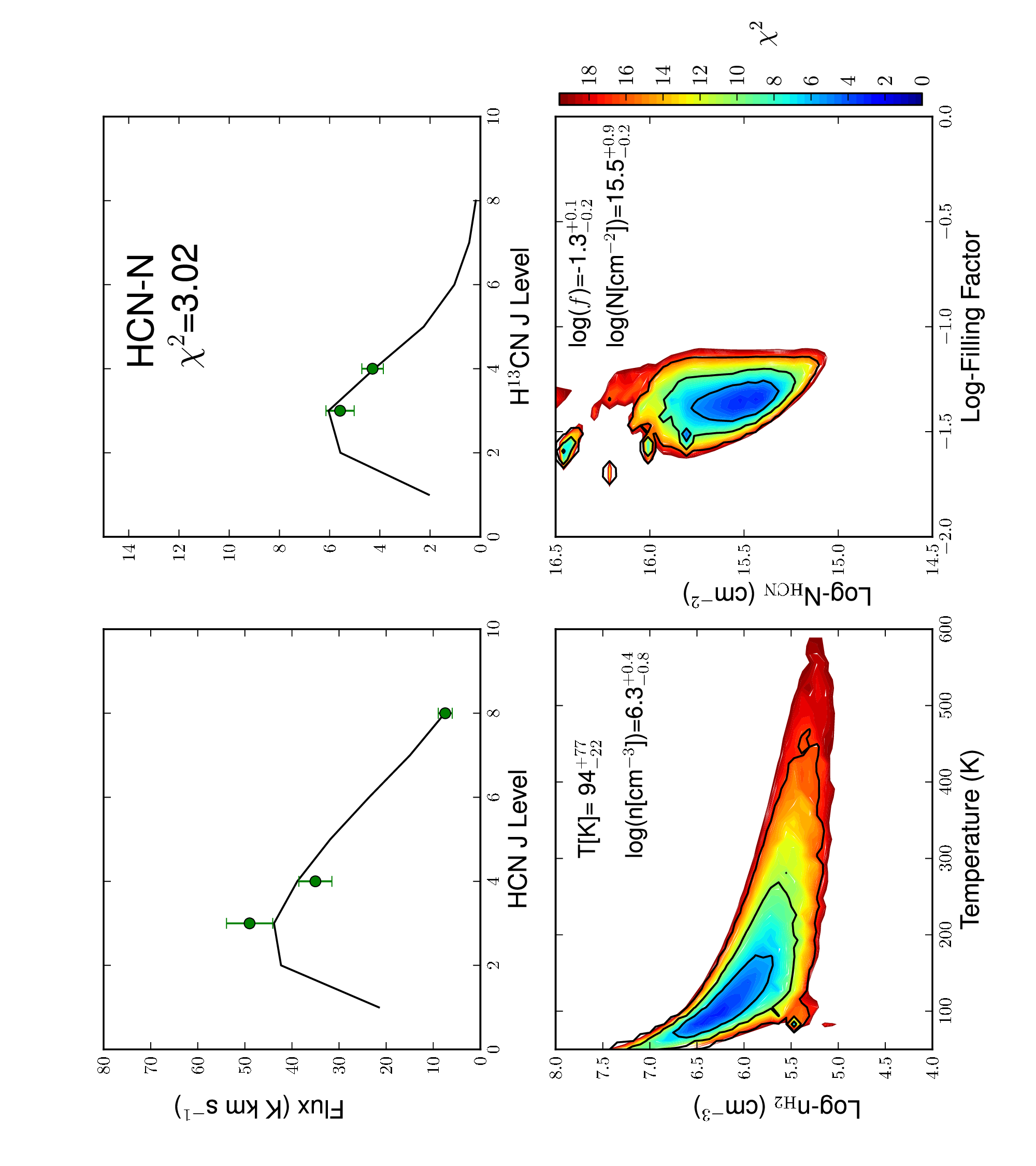}
\includegraphics[scale=0.4, angle=270]{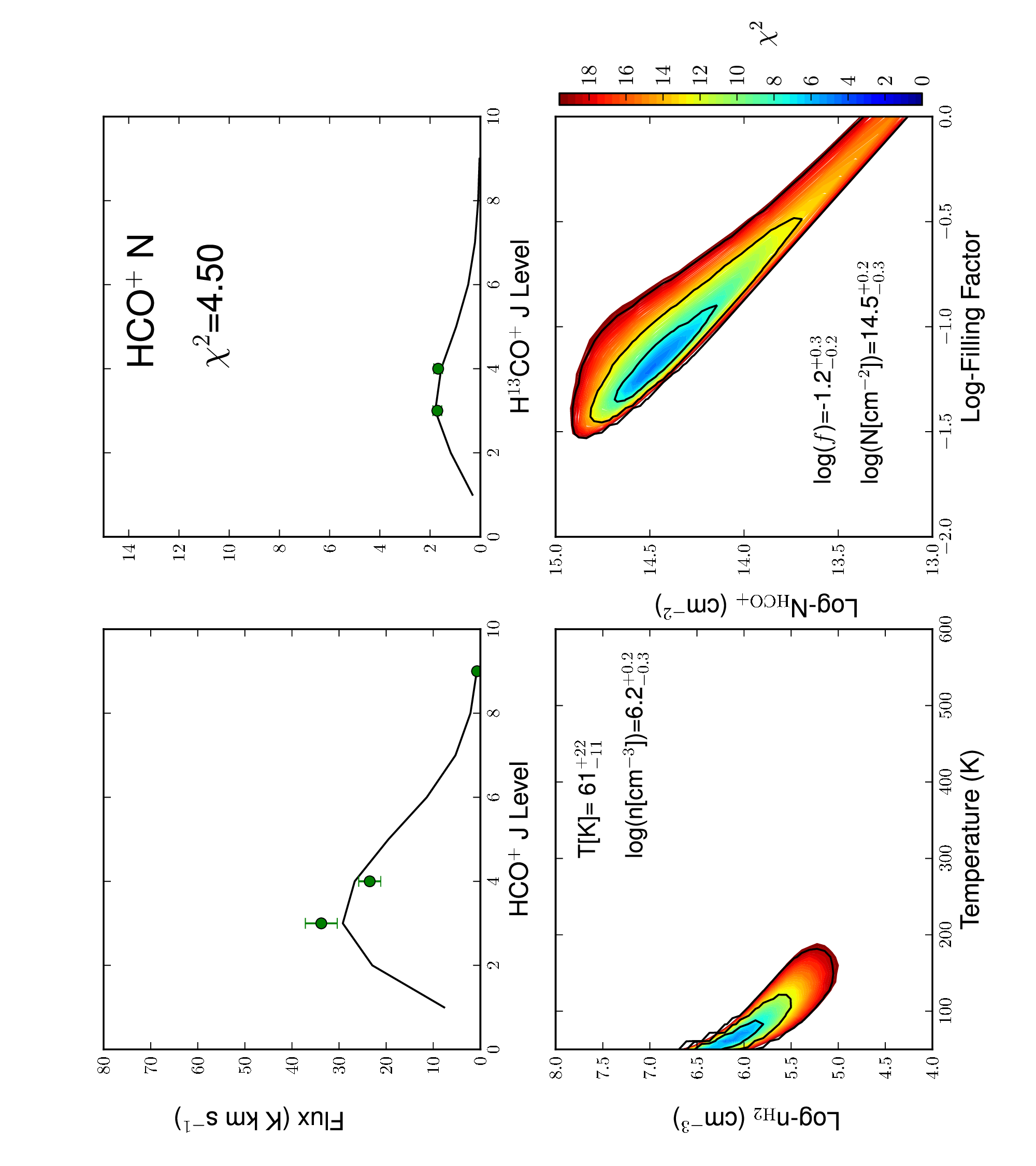}
\caption{Chi-squared fits to a 1-component model of HCN and \hco\, excitation for feature N in the CND. Rows are the same as for Figure \ref{hcn_north}.  }
\label{hcn_N}
\end{figure*}

\begin{figure*}[tbh]
\hspace{-0.5cm}
\includegraphics[scale=0.4, angle=270]{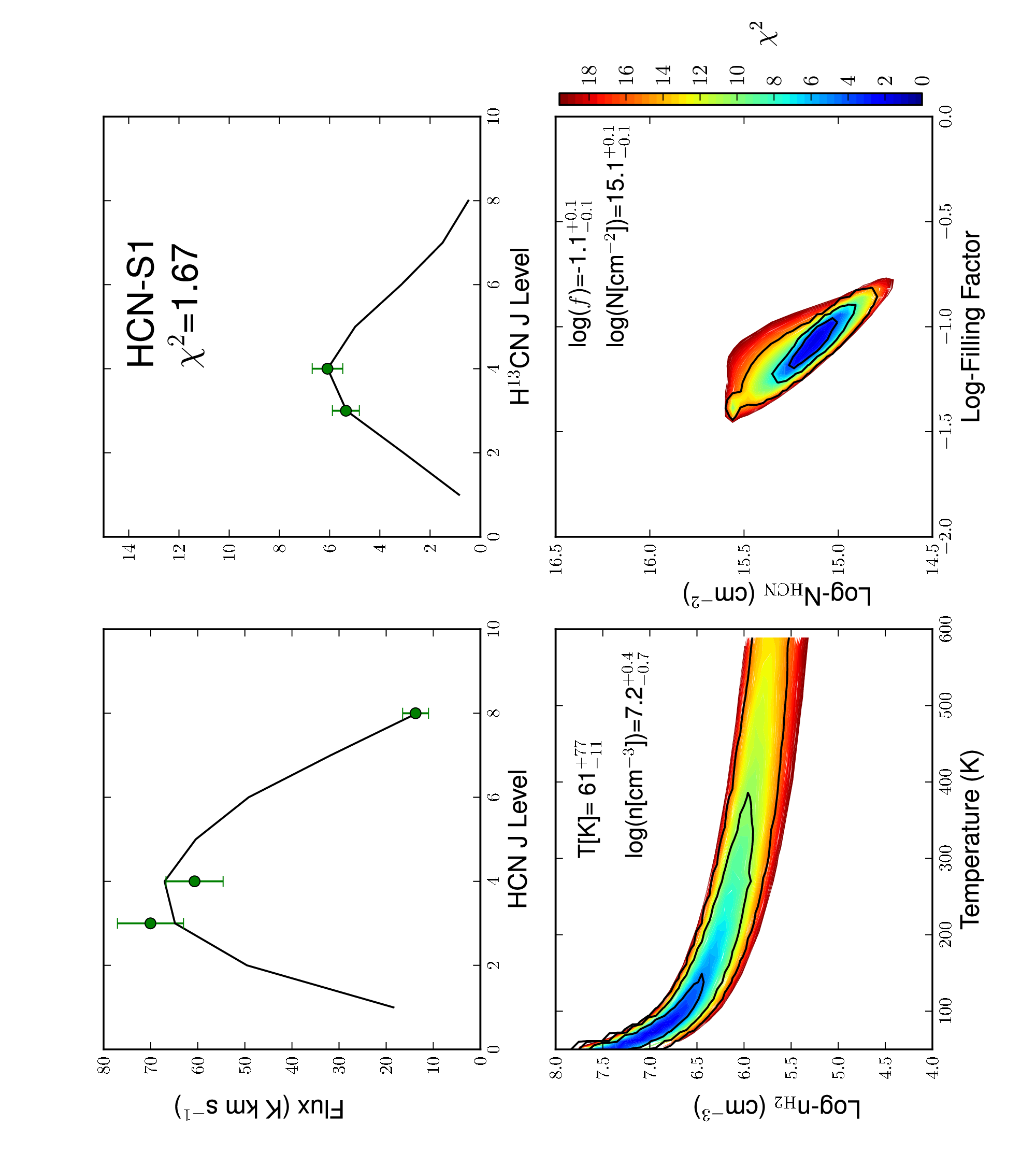}
\includegraphics[scale=0.4, angle=270]{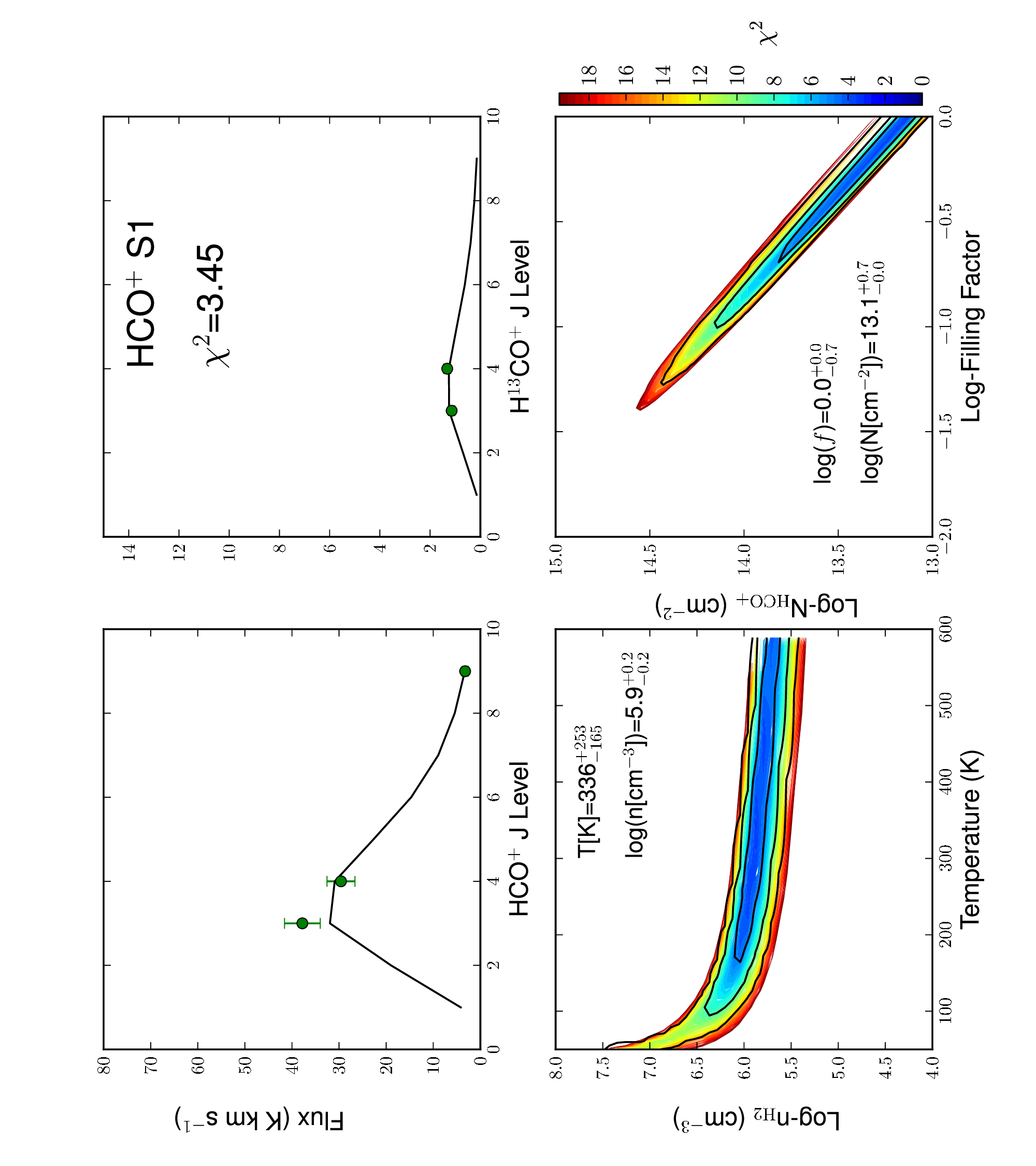}
\caption{Chi-squared fits to a 1-component model of HCN and \hco\, excitation toward feature S1 in the CND. Rows are the same as for Figure \ref{hcn_north}.}
\label{hcn_S}
\end{figure*}

\begin{figure*}[tbh]
\hspace{-0.5cm}
\includegraphics[scale=0.4, angle=270]{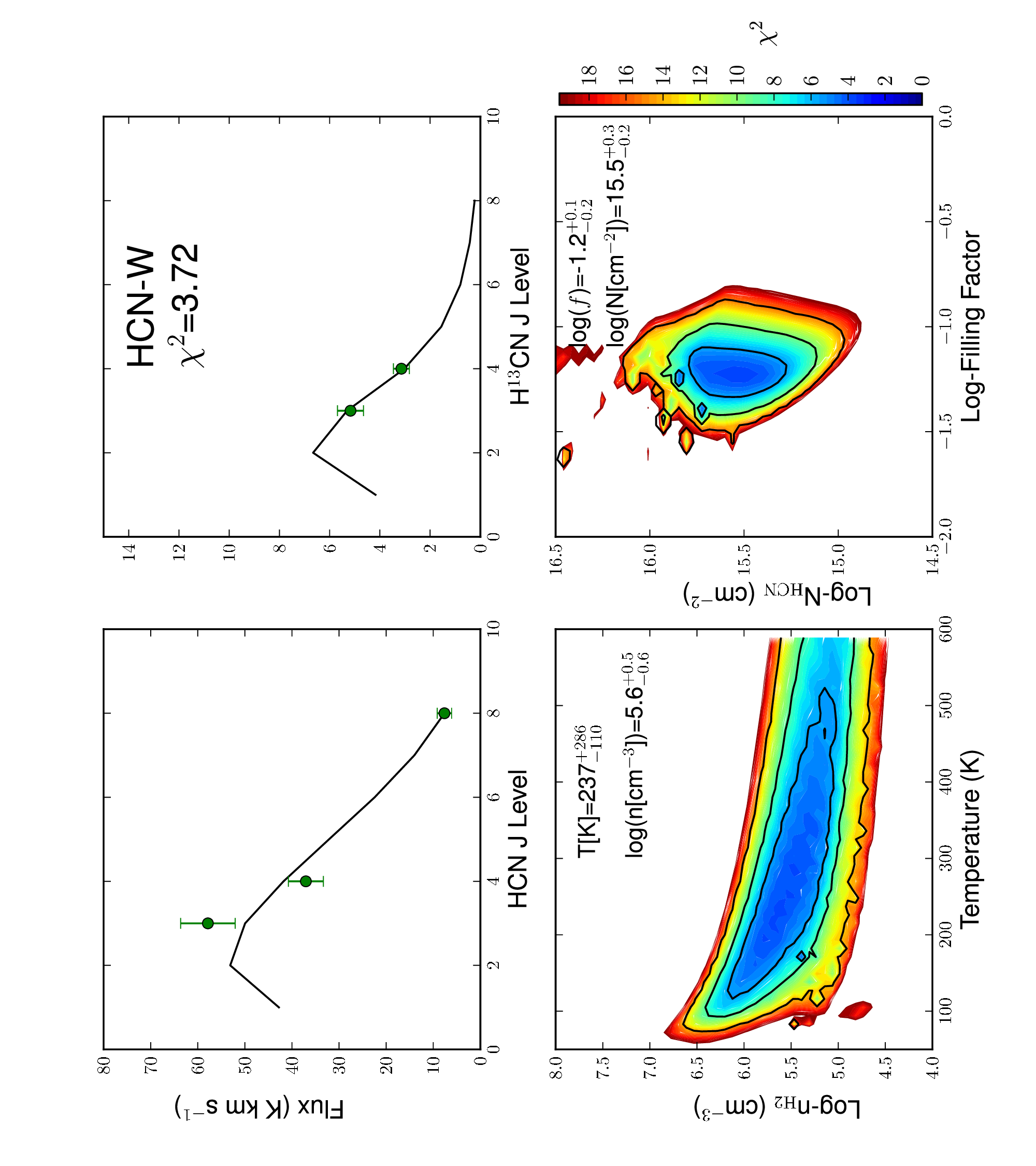}
\includegraphics[scale=0.4, angle=270]{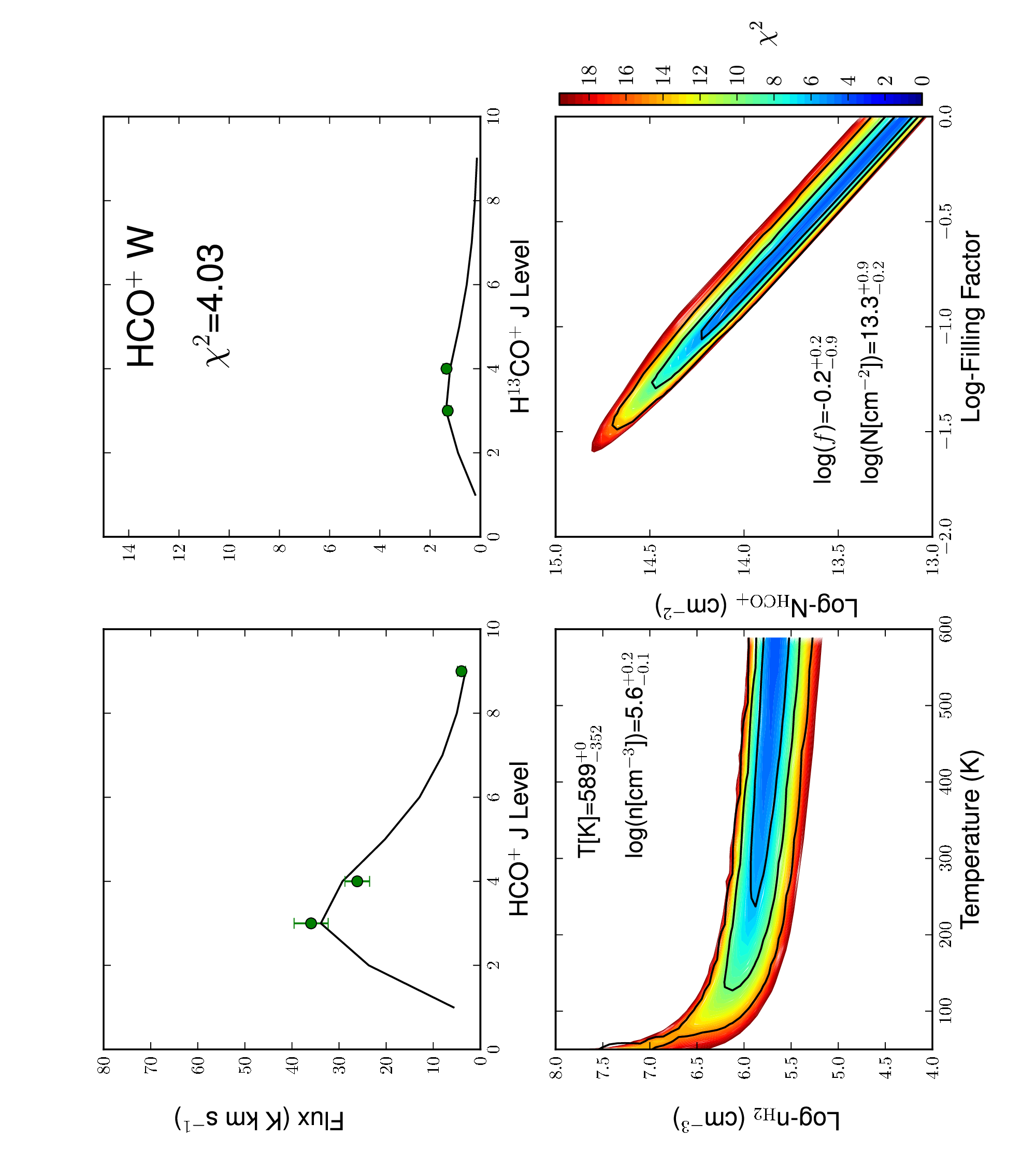}
\caption{Chi-squared fits to a 1-component model of HCN and \hco\, excitation toward feature W in the CND. Rows are the same as for Figure \ref{hcn_north}.}
\label{hcn_W}
\end{figure*}

\begin{figure*}[tbh]
\hspace{-0.5cm}
\includegraphics[scale=0.4, angle=270]{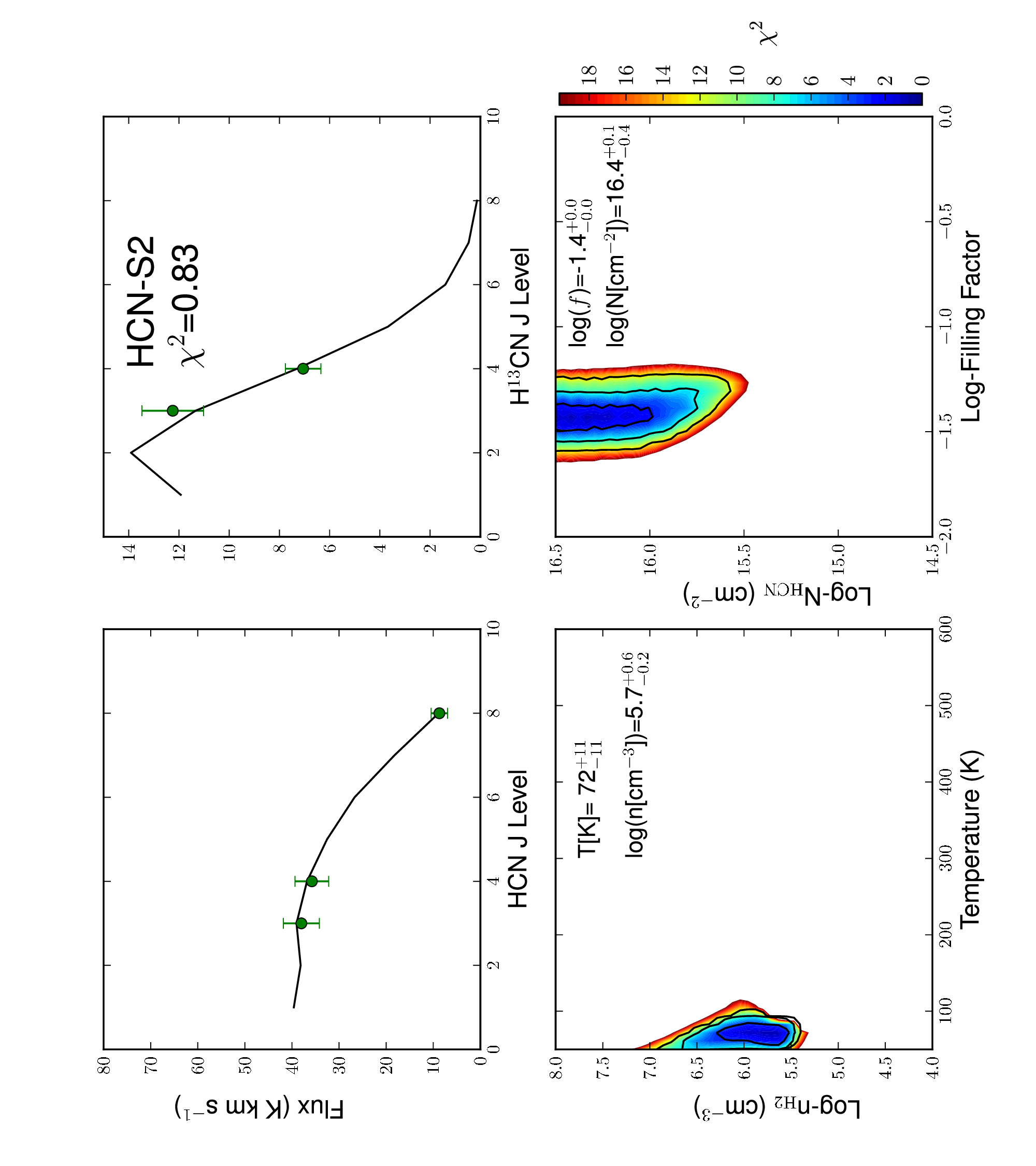}
\includegraphics[scale=0.4, angle=270]{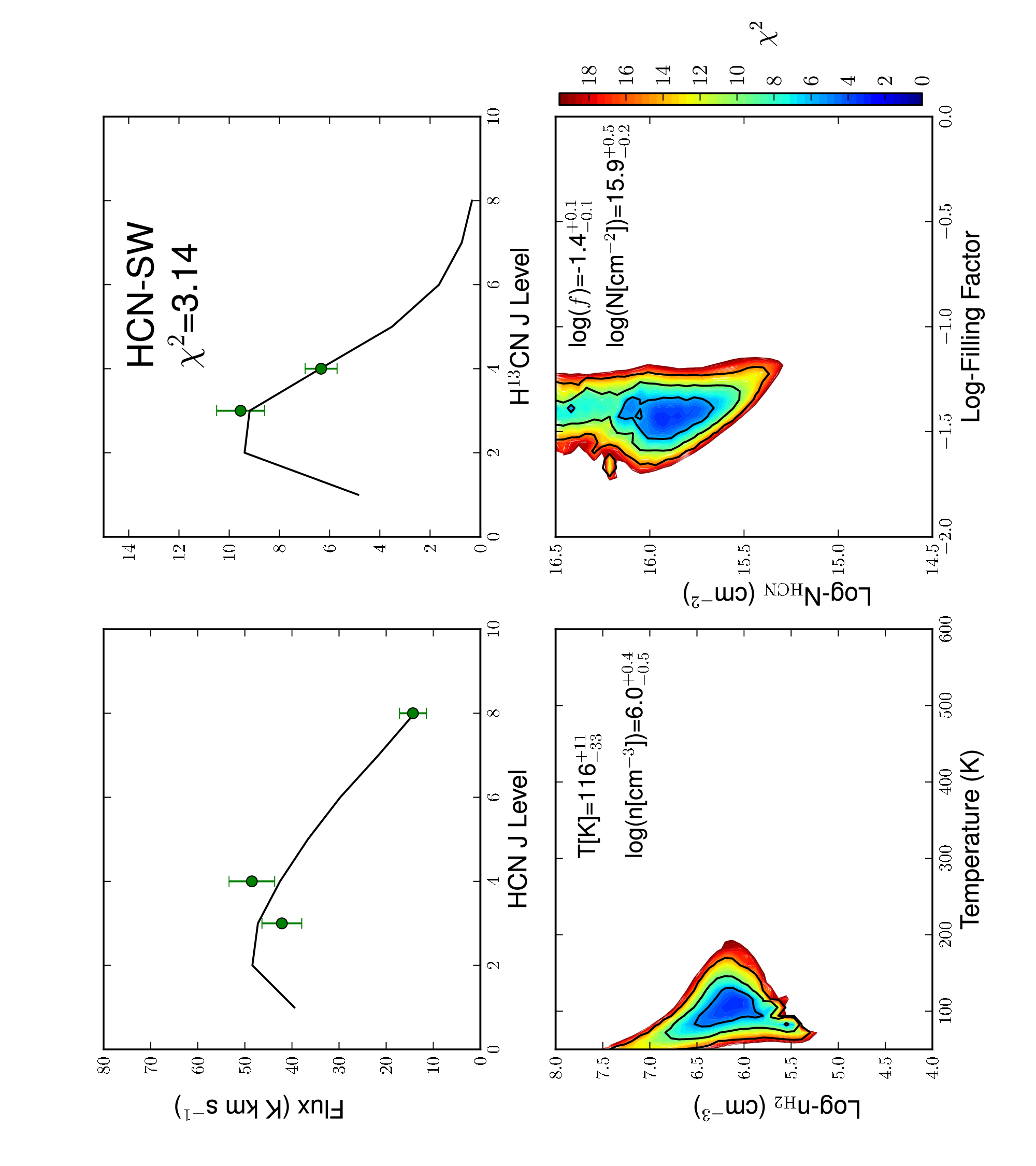}
\caption{Chi-squared fits to a 1-component model of HCN excitation toward feature S2/SW in the CND. Rows are the same as for Figure \ref{hcn_north}.}
\label{hcn_SW}
\end{figure*}

	 	In addition to fitting for the brightness temperatures integrated over the entire line profile, we also fit to the brightness temperatures integrated over limited ($\Delta$v = 20 \kms\hspace{-0.12cm}) velocity ranges corresponding to individual clumps identified in interferometric studies of the CND. For the five features we analyze, we find reasonable fits to observations of all features except for the \hco\, observations of the S2/SW clump. Plots of the $\chi^2$ distribution for all features are shown in Figures \ref{hcn_N}, \ref{hcn_S}, \ref{hcn_W}, and  \ref{hcn_SW}, and the fit parameters are reported in Table \ref{Results}. Within our uncertainties, the volume densities of individual features range from $n=10^{5.0}$ to $10^{7.6}$ \cm\, (for feature S1).  The densities and temperatures derived from fits to HCN and \hco\, are consistent except in the case of feature S1, where the HCN-derived density is higher by an order of magnitude. In this case, the HCN fits favor a cooler temperature ($\sim$ 100 K) than fits to \hco\, which favor a temperature $>170$ K.  For \hco\, fits to S1, and fits for both molecules to feature W, the constraints from the observed lines are insufficient to constrain the temperatures, and acceptable fits are found for temperatures up to 600 K, the largest value considered in our grids of parameter values. In general however, fits to the limited velocity ranges of individual features yield lower temperatures and higher densities than fits to the entire profile. 
	
\begin{figure*}[tbh]
\hspace{-0.5cm}
\includegraphics[scale=0.5, angle=270]{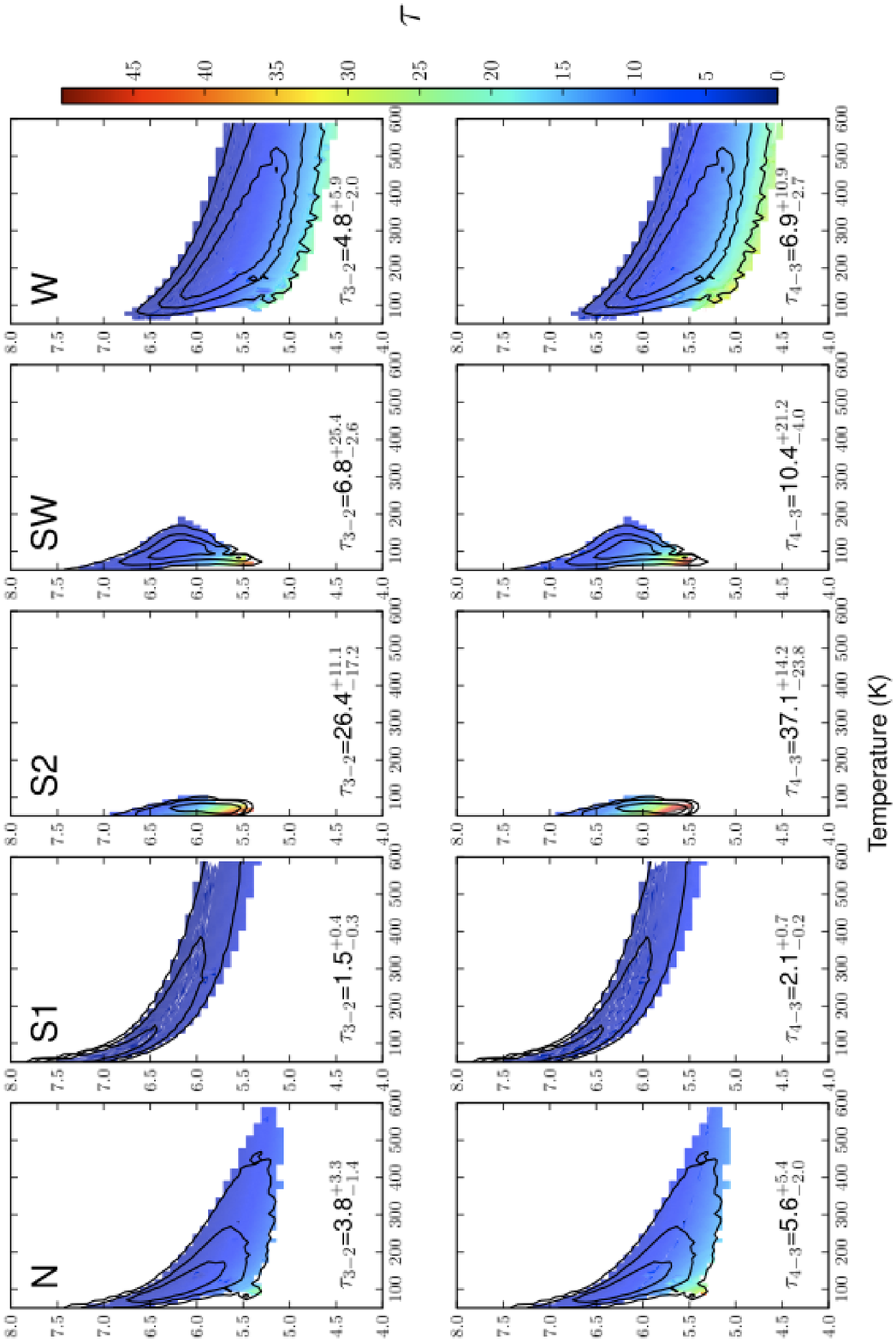}
\\
\\
\caption{HCN line opacities for the 3-2 (Top) and 4-3 (Bottom) transitions, derived from fits to individual features. Contours show the 1-,2-, and 3-$\sigma$ deviations from the most likely temperature and density over the full grid of physical conditions that were considered.}
\label{hcn_OP}
\end{figure*}

	 	Similar to the results from fits to the majority of the line profile, the \hco\, filling factors from fits to individual features are uniformly larger than the HCN filling factors. Possibly related to this, but more likely a factor of the lower relative \hco\, abundance in the CND, the emission from \hco\, features is generally optically thin, with only emission from feature N being optically thick. In contrast HCN emission is extremely optically thick, with opacities ranging from 1.5 to potentially higher than 30 (for feature S2, if T $<$ 100 K and n $< 10^6$ \cm).  The distribution of opacities derived from our fits to individual features are shown in Figures \ref{hcn_OP} and \ref{hco_OP_full}. For features N, S1, and W, the HCN optical depths are lower than those derived from fits to the majority of the line profile. 
		
		\subsubsection{Feature properties} 
		If the source being observed is smaller than the telescope beam, its brightness temperature will be diluted. Assuming the features we observe are circularly symmetric Gaussian clumps at the center of the circular beam, the filling factor which describes this dilution is given by the expression:
		
	\begin{equation} f = \frac{\theta_{clump}^2}{\theta_{MB}^2+\theta_{clump}^2},
	\label{ff}
	\end{equation}			
	
where $\theta_{MB}$ is the FWHM of the telescope beam and $\theta_{clump}$ is the FWHM of the clump. Taking the clump to be a sphere with a radius equal to half of the FWHM, we use the best fit filling factors to determine equivalent radii for the observed features. These radii range from 0.1 to 0.15 pc, and are reported in Table \ref{Results}. In all cases, the radii we derive are slightly smaller than those derived from interferometric observations of the individual clumps \citep{MCHH09}. 

Using the clump radii we derive, we can also estimate masses for individual clumps, assuming a uniform clump density equal to the best-fit value. The resulting masses are reported in Table \ref{Results}. The typical clump mass is a few hundred solar masses, except for clump S1, for which we estimate a mass of a few $10^4$ M$_{\odot}$ from our HCN observations. All of the masses derived using HCN are at least an order of magnitude lower than the virial masses determined for these clumps by \cite{MCHH09} using HCN 4--3. For \hco\, we only have model fits for three features, and two of these features have large filling factors inconsistent with emission from a clump. For the remaining feature (N), the mass we derive using \hco\, ($\sim$ 840 M$_{\odot}$) is similar to that derived for this clump using HCN, and consistent within the uncertainties with the mass derived for this feature by \cite{MCHH09}.

Finally, given the clump radii we derive, we can also compare the observed line brightness temperatures from our convolved data to those in the unconvolved HCN 8--7 data, which has a higher resolution (8.9$''$). We find that, for the derived clump radii, the difference in beam dilution should lead the unconvolved HCN 8--7 brightness temperatures to be a factor of 3-4 times larger than the HCN 8--7 brightness temperatures from data convolved to the resolution of the HCN 3--2 beam (23.6$''$). However, we find that it is only typically a factor of 2-3 times brighter for these features. This suggests that the sources are actually slightly more extended than our analysis suggests, and/or that other estimates made when deriving radii from the filling factors (that the sources are circular, Gaussian, and lie in the center of the beam) may not hold.

	 	Overall, our results indicate that conditions in the individual clumps we examine are consistent with the average temperatures and densities in the bulk of the CND gas, although some features, such as S1, may be slightly denser. However, the results presented here for individual features are based upon line intensities which still likely suffer from the superposition of line profiles from multiple clumps in the large beam of these observations. Higher spatial-resolution observations are necessary in order to fully isolate the contribution to the CND emission from these individual clumps, and to properly model their line profiles. 

		\subsection{Radiative excitation of HCN due to the 14$\mu$m background field}		
		Thus far in these analyses, we have been assuming that HCN (and \hco) are purely collisionally excited. However, our detection of the $J$ = 4-3 $v_2=1$ line of HCN suggests that the excitation of HCN in the CND is not entirely collisional, and that radiative excitation also plays a role. There are two possibilities for this radiative excitation, either the gas traced by HCN is irradiated externally by a mid-infrared background field, or it is irradiated internally, through embedded hot dust mixed with the gas. We use the mid-infrared background spectra measured by \cite{Lutz} to run multiple RADEX models incorporating a mid-infrared background radiation field, and find that the results from these runs are indistinguishable from the case in which no mid-infrared background radiation is included, for the full range of temperatures and densities we consider. We thus conclude that background 14 $\mu$m radiation fields up to several times the highest value measured toward the central parsec ( 250 Jy nsr$^{-1}$, for an aperture centered on Sgr A*) are a negligible contribution to the excitation of HCN. This favors embedded hot dust as the most likely excitation mechanism for the observed $J$ = 4-3 $v_2=1$ HCN line, which we explore further with excitation analyses using RATRAN.  

	\section{RATRAN}
		\label{rat}
		\subsection{Input Parameters}
		We model the radiative excitation of HCN in the CND with RATRAN, using the same collisional and radiative excitation data as for the RADEX analysis. Like RADEX, RATRAN is a one-dimensional non-LTE code. However, RATRAN differs from RADEX in that it uses Monte Carlo techniques to more carefully sample the radiation field, combined with an accelerated lambda convergence method which allows the code to operate efficiently even for extremely high opacities. Unlike RADEX, RATRAN is not limited to the assumption of uniform source conditions, but can model gradients in the physical parameters, allowing for a more realistic modeling of the shape of the emergent line profiles, as opposed to the simple RADEX model of a rectangular line profile (having constant intensity and opacity). For simplicity, we assume a spherically-symmetric clump, with a uniform HCN abundance. 
		
		RATRAN modeling requires several more constraints on the gas properties. We assume the same turbulent line FWHM (20 \kms\hspace{-0.1cm}) as used in RADEX fitting. However, we now also define a separate velocity gradient, dv/dr, for the CND, which we assume to be 150 \kms\, pc$^{-1}$, consistent with observations of the kinematics of CND gas \citep{Chris05,Martin12}. For the source radii, we adopt a value taken from interferometric observations (assuming spherical clumps): $\sim$0.19 pc, consistent with our best-fit filling factors from the RADEX analysis. We also have to input the abundance of HCN compared to that of its primary collision partner (H$_2$). We assume the radiative excitation of HCN is predominantly due to emission from hot dust which is mixed with the gas, and fix the intensity of the internal mid-infrared radiation field by assuming isotropically-distributed dust with temperatures between 100 to 150 K, a gas-to-dust ratio of 100, and assuming a dust emissivity model \citep[grains with no ice mantle and $\sim10^5$ years of coagulation,][]{Ossenkopf}. For typical CND column densities, this results in emission which is optically-thick at the wavelengths of the 14 $\mu$m rovibrational transitions. As there are now more free or uncertain parameters to our fit, including the dust temperature and the HCN abundance, we do not try, as with RADEX, to produce grids of conditions to constrain the most likely solution. Instead, we look for the existence of well-fitting solutions, acknowledging that they may not be unique. 

		\subsection{Overlap of the 14 $\mu$m rovibrational lines}
		\label{rex}
In addition to taking into account radiative excitation due to embedded hot dust, we also consider the effects of line overlap in the Q-branch transitions at 14 $\mu$m, which connect the $v_2=0$ and $v_2=1$ states. 

If the opacity at the wavelength of the $v_2$ = 1--0 rovibrational transitions ($\lambda = 12.5-16.0$ $\mu$m) is sufficiently high, then the mid-infrared photons at this wavelength are less likely to escape the vicinity of a given gas molecule, and can instead be re-absorbed by the molecule. This trapping phenomenon reduces both the critical density for collisional excitation and the infrared background necessary for radiative excitation of the $v_2=1$ line. 
		
\begin{figure*}[tbh]
\includegraphics[scale=0.4]{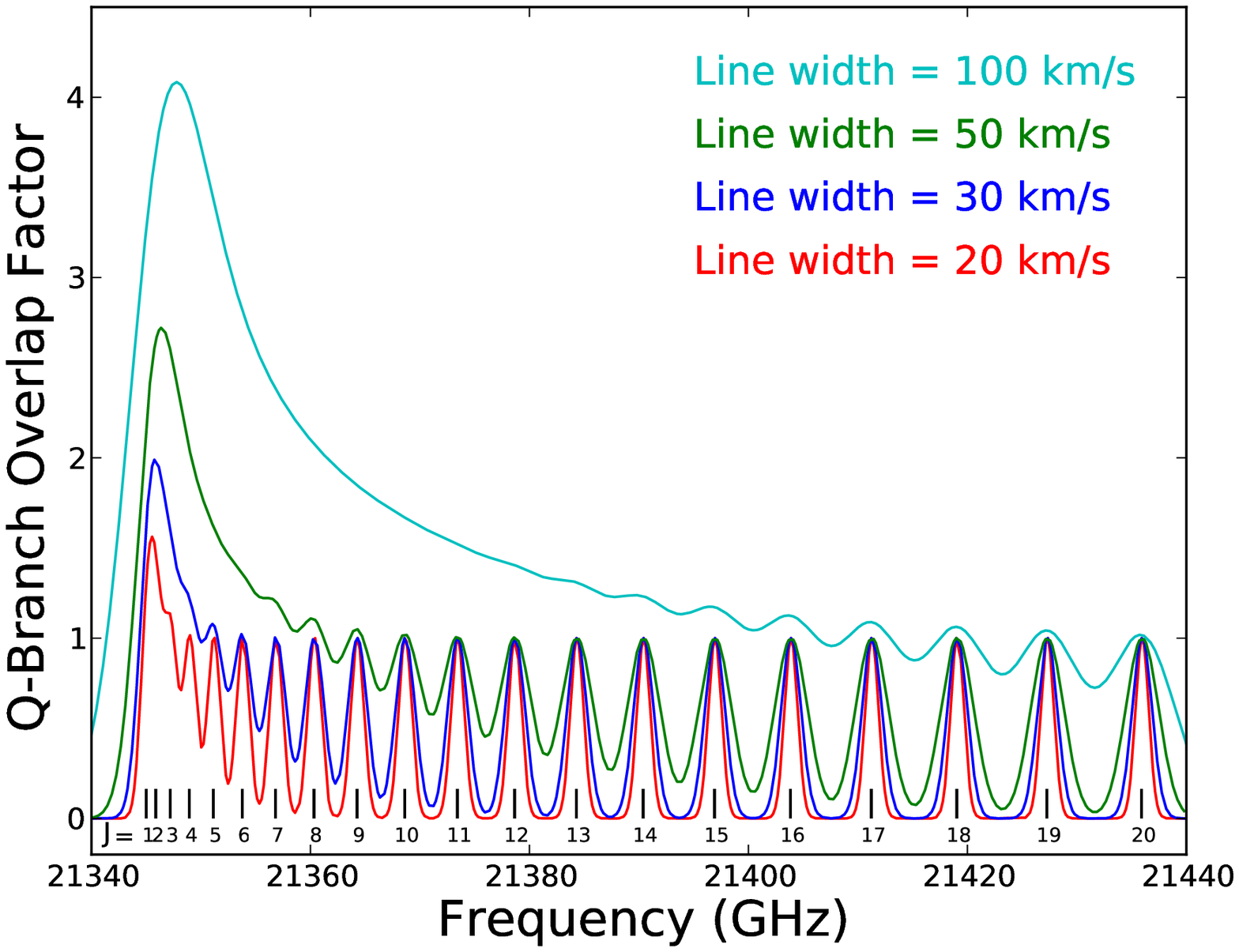}
\caption{The degree of overlap in the Q-branch rovibrational lines of HCN for a range of line widths. An overlap factor of 1 signifies no overlap.}
\label{Qbranch}
\end{figure*}

		Although both RATRAN (and RADEX) take into account the trapping of radiation for high opacities, neither code treats the effects of overlapping line emission on the excitation. The 14$\mu$ m rovibrational transitions of HCN occur in three bands: the R-branch ($\Delta$$J$ = +1), Q-branch ($\Delta$$J$ = 0), and P-branch ($\Delta$$J$ = -1). The Q-branch transitions in particular are sufficiently closely spaced that for CND line widths (20 to 100 \kms\hspace{-0.1cm}) transitions up to $J \sim$ 4 will overlap. We calculate the expected overlap as a function of line width and make a correction for this overlap (as seen in Figure \ref{Qbranch}, lines overlap by a factor of 1.5 to 2.75 for $J$ up to 5) by dividing the Einstein-A values of the Q-branch transitions by the overlap factor, which will increase the populations in the $v_2=1$ states. Here, the overlap factor for each line is defined as the contributions from Gaussian profiles of all the lines (where the peak intensity of the Gaussian is normalized to 1) summed at the frequency of each line. 
				
	We find that including the effects of Q-branch overlap for line widths as large as 100 \kms\, has a substantial effect on the predicted intensity of the $v_2$=1 J = 4-3 line, while the effect of overlap on the predicted intensities of the $v_2$=0 rotational lines is much smaller.  Accounting for this overlap alters the $v_2=1$ $J$=4--3 line intensity by a factor of 2-4 for line widths of 50-100 \kms, however the $v_2=0$ line intensities vary by $<10$\%.  The smaller variation observed in the $v_2=0$ lines is likely because the Q-branch transitions are $\Delta$$J$ = 0 transitions, and so contribute less to altering the level populations of the $v_2=0$ rotational lines than the R-branch transitions. The effect of Q-branch overlap on the $v_2$=1 ladder is stronger because of the increase of the effective lifetime of the $v_2$=1 upper states, giving a greater probability for the rotational transition to occur before the vibrational decay.  For the following analysis, we include the effects of overlap for the HCN 4--3 line widths from interferometric observations of individual clumps: 50 \kms\, for the South pointing, and 100 \kms\, for the Southwest pointing \citep{MCHH09}.
	
		\subsection{ Radiative excitation of HCN by embedded dust}
		\label{ratran}
	The $J$ = 4--3 $v_2=1$ line is predominantly detected at velocities around -20 \kms\hspace{-0.1cm}, corresponding to the Southwest pointing (or the S2/SW clump).  For this source, we adopt an interferometrically-determined radius of 0.25 pc  from \cite{MCHH09}. We then run a coarse grid of RATRAN models covering the range of conditions indicated by our RADEX fitting to the observed line intensities toward Southwest (T=100 K, n = $10^{5.5}-10^{6.4}$ \cm, [HCN/H$_2$] = $5\times10^{-9}- 1\times10^{-8}$), varying the dust temperatures between 100 and 175 K.  We find that dust temperatures of $\sim125-150$ K are required to generate the observed $J$ = 4--3 $v_2=1$ line strength ($\int$ T$_{MB}$ dv = 2.28 K \kms). The T$_{vib}$ for these models are $\sim$150 K, smaller than the T$_{vib}$ we calculate from the simplified analysis underlying Equation 2 ($\sim200$ K, although this value could be an overestimate if the H$^{13}$CN 4--3 line used to calculate this value is actually optically thick). 
	
	These dust temperatures are also sufficient to cause the derived density for South-1 to be up to a factor of 5 lower than predicted by RADEX. Thus, as we discuss further in Section \ref{dd}, if this hot dust is widespread throughout the CND and not just localized in feature S2/SW, this could bring the densities derived using HCN for the southern emission peak into agreement with those derived by RT12 using CO. If our density estimates for other clumps are indeed overestimated due to the radiative excitation of HCN,  $v_2$ =1 emission should be seen in these clumps as well. It should be noted that these results assume a homogenous, spherical source. If there are spatial variations in temperature, density, or abundance, these would alter our assumption that all the line emission from the $v_2=0$ transitions is arising from the same gas as the $v_2=1$ transition, and could change the inferred dust temperature, and thus the degree to which radiative excitation could affect the derived densities. 

\section{Discussion}
\label{Dis}	
		\subsection {Hot dust in the CND}

The detection of the $J$ = 4-3 $v_2=1$ HCN line  toward the southwest emission peak of the CND and subsequent RATRAN modeling of its intensity indicate that HCN emission from this region is radiatively excited, and that a 125-150 K dust component is necessary to explain the observed strength of this line.

However, existing observations have not detected this hot dust component in the CND. \cite{Etx11} fit the far-infrared and submillimeter emission spectrum of this region with photometric data from 21.3 to 180 $\mu$m, indicating the presence of at least three temperature components in the dust: 23, 44.5, and 90 K. More recently, Lau et al. (2013) used the FORCAST instrument on the Stratospheric Observatory For Infrared Astronomy (SOFIA) to map emission from the CND and central parsec at 19.7, 31.5, and 37.1 $\mu$m and to construct a map of the dust color temperature, showing clearly that the hottest dust, with temperatures up to 150 K, originates in the central cavity, while in the CND, the dust color temperature ranges from 60-90 K, consistent with lower-resolution observations at similar wavelengths by \cite{Telesco}.

We consider several potential mechanisms for heating CND dust to temperatures of 125-150 K. \cite{Lau13} find that, given the radiation field from the central cluster, and assuming a distance of 1.4 pc and dust grain sizes of 0.1 $\mu$m, the equilibrium CND dust temperature should be $\sim$ 90 K. Assuming a smaller distance (1 pc, the projected distance of feature SW from the central star cluster) will only increase the assumed dust temperature by 10\%. However, if there are smaller dust grains present in the CND (for example, with an order-of-magnitude smaller size  of 0.01 $\mu$m), then approximating the dust temperature to depend on particle size $a$ as $a^{1/6}$ \citep{Krugel03},  the equilibrium temperature of those grains could exceed 125 K. However this is approaching the regime of small grains, where this approximation breaks down.

In sufficiently dense regions, the dust could also be heated via collisions with the gas. Gas-dust thermal coupling should be effective at typical densities of $\sim10^5$ \cm\, \citep{Goldsmith01,Juvela11}, although the density required is increased if the cosmic ray background is higher than typical interstellar values \citep{Clark13}. As noted in the introductory section of this paper, gas temperatures in the CND have been found to range from 50-400 K, and exceed 200 K in the atomic gas on the inner edge, so given that we constrain the typical CND densities to be between $10^5$ and a few $10^6$ \cm, it is not unreasonable to expect the dust and gas should be coupled, with some fraction of both having temperatures $>125-150$ K.  We thus find that either a population of small dust grains or heating via coupling with the gas are potentially viable explanations for the hot dust component we infer from these observations, and should be investigated further. We additionally discuss potential heating sources for the gas in Section \ref{env}

There are several possible reasons why such a hot dust component in the CND could have been missed by previous observations. First, the wavelengths observed by \cite{Etx11} and \cite{Lau13} are slightly longer than the wavelengths of the HCN rovibrational transitions, so they do not constrain the existence of hotter dust components which could dominate the dust emission at the wavelengths of the rovibrational lines of HCN. For example, a 12.5-20.3 $\mu$m color temperature map of the `minispiral' region interior to the CND indicates dust color temperatures ranging from 200 to 270 K \citep{Cotera99b},  significantly higher than the $\sim$ 150 K dust temperatures derived by Lau et al. and others for the minispiral.  Also, short wavelength emission from the hottest dust should originate from the inner edge of the CND, and given the orientation of the CND, this emission from clump S2/SW should lie on the far side of the bulk of the gas in the southern emission peak of the CND (see, e.g., Lau et al.). Emission from hot dust in this source could then be significantly extinguished. If there are significant quantities of hot dust all along the inner edge of the CND, we predict that this hot dust component could be indirectly detected via the radiative excitation of HCN: the 14 $\mu$m rovibrational lines should be seen in absorption against the background dust continuum (where they are less extinguished by the dense CND torus), and emission from the $v_2=1$ rotational lines of HCN should also be present in other high-column density CND clumps lying along its inner edge. 

We also note that it is not clear whether radiative excitation may also contribute to the excitation of \hco. We do not detect the $v_2=1$ lines of \hco, however this may not rule out radiative excitation, as \hco\, has a lower abundance than HCN in the CND, as we discuss in Section \ref{env}, making these lines too faint to be detected in our data. However, the rovibrational transitions of \hco\, also occur at 12 $\mu$m, at which wavelength archival ISO spectra show the mid-infrared emission from the CND to be a factor of 2-3 less intense than at 14 $\mu$m, so it is possible that radiative excitation is not important for \hco. 

In contrast, HNC should be more easily radiatively excited than HCN or \hco, as the rovibrational transitions of its $v2=1$ bending mode occur at 21.5 $\mu$m, where the background dust emission is stronger. For gas densities $\sim10^6$ \cm, HNC can be radiatively pumped by embedded dust with temperatures as low as 85 K \citep{Aalt07a}.  As a result, HNC should be more sensitive to the presence of warm dust in the CND than HCN. Future observations of the HNC/HCN ratio and a search for the $v_2=1$ rotation-vibration lines of HNC would then provide a useful diagnostic of the distribution of warm dust in the CND, and its resulting influence on the radiative excitation of molecules in the CND. 

		\subsection{CND Densities}				
		\label{dd}
		
		 Comparing our RADEX-derived HCN and \hco\, densities to those derived by RT12 using CO, we find that in general, the best-fit densities from our fits to the  majority of the line profile are consistent within the uncertainties with densities from RT12 toward the northern and southern emission peaks of the CND, although they tend to be slightly higher.  However, the best-fit density for the South-1 pointing determined from our HCN observations, (log [n (\cm)] =  6.5$^{+0.5}_{-0.7}$ ), is higher than the CO-derived densities found by RT12 for the southern emission peak (log [n (\cm)] =  5.2$^{+0.4}_{-0.2}$), and does not agree within the uncertainties of both measurements. There is also some tension in the best-fit temperatures;  the temperature determined from HCN line intensities toward South-1 (as well as the temperature determined from \hco\, toward the North pointing) is lower than allowed by the RT12 CO fits to the high-density component of gas.  There are two main scenarios which can explain this discrepancy: either there truly is higher density (and cooler) gas in the southern emission peak of the CND, or (as is the case for the separate Southwest peak), radiative excitation contributes substantially to the excitation of the HCN lines, with the resulting alteration in the level populations mimicking a higher density than is actually present. 
				
			It is possible that HCN (and \hco\,) trace gas which is preferentially excited in higher-density clumps of the CND, as the critical densities of the highest lines we measure (HCN 8--7 and \hco\, 9--8) are $\sim$ $10^9$ \cm, and $\sim$ $10^8$ \cm, respectively, whereas the critical density of the highest line measured by RT12, CO 16--15, has a critical density of just a few times $10^6$ \cm, three orders of magnitude lower. Our fits to limited velocity ranges corresponding to individual interferometrically-detected clumps also tend to slightly indicate higher densities than fits to the entire line profile, which suggests that there may be denser clumps embedded in the CND, even though these higher densities may not be typical of conditions in the bulk of the CND gas, as traced by CO. However, if radiative excitation is important outside of clump S2/SW (where we detect the $v_2=1$ J=4-3 line of HCN), and there is also a hot dust component in the southern emission peak, this could also lower the gas density we derive for this source, potentially bringing it into agreement with the lower density measured by RT12. As we do not detect $v_2=1$ J=4-3 emission toward the CND's southern emission peak, our current observations are not sufficient to distinguish between these scenarios; sensitive observations of the $v_2=1$ lines of HCN toward more positions in the CND should be conducted in order to determine whether the radiation from hot dust plays a role in the observed excitation of HCN throughout the CND. 
			
		\subsubsection{Is CND gas virialized?}
		Previous interferometric observations of the CND with HCN and \hco\, have  suggested that, if virialized, the CND gas should have extremely high densities, on the order of $10^7-10^8$ \cm \citep{Chris05,MCHH09}. However, even the highest density we find (feature S1) is substantially lower than the large virial densities calculated by \citeauthor{MCHH09}, which are on the order of a few times $10^8$ \cm. In general, we find densities less than a few $10^6$ \cm, suggesting that the clump densities determined by \citeauthor{MCHH09},  \cite{Shukla04} and \cite{Chris05} are overestimated, and CND clumps are not in virial equilibrium. We also compare our derived densities to the Roche limit for stability in the CND, which is $\sim$ 10$^7$ \cm. The densities we derive for all positions in the CND except one are lower than this value, meaning that the majority of the gas in the CND is not tidally stable. The densities allowed by our single-component HCN fit to feature S1 are  log [n (\cm)] = $7.2^{+0.4}_{-0.7}$, and so for this clump, we cannot rule out the possibility that this clump could be marginally stable against tidal disruption. Intriguingly, this is also the only CND clump that has a strong submillimeter counterpart \citep{Liu13}. However, overall our analysis indicates that, consistent with the findings of RT12, the bulk, if not the entirety, of the CND gas is not tidally stable, and the observed clumps must be transient features. 
		
		\subsection{Feature SW/S2}
		\label{SWfit}
		The most unusual properties from this analysis are associated with feature SW/S2, at a velocity of -20 \kms\hspace{-0.1cm}. This is the feature from which we detect the $J$ = 4--3 $v_2=1$ line of HCN, as well as the only feature from which we see significant self absorption in the J= 3--2 and 4--3 spectra of both HCN and \hco\, (Figures \ref{HCN}, \ref{HCO}). 
		
		RADEX fits to this feature indicate that it has the highest column density of all features we survey (the best-fit HCN columns are $10^{16.4}$ cm$^{-2}$ and $10^{15.9}$ cm$^{-2}$ toward SW and S2 respectively), and that extremely high opacities ($\tau \sim 10 - 50$ ) are needed to account for the observed self absorption in the line profiles. However, this self absorption could also be explained by multiple excitation components along the line of sight. In these scenarios, the opacities (and thus column densities) could be lower than those found by our RADEX fits. One way to test this is by looking for emission from rarer isotopologues of HCN and other species. Although the opacities from our RADEX fits are sufficiently high that the HC$^{15}$N lines should be detectable \citep[with intrinsic line intensities of a few hundred mK, assuming a Galactic center $^{15}$N/$^{14}$N ratio of 600][]{WR94}, the large beam sizes of our observations dilute the expected signal so that it is still lower than our observed upper limits for the HC$^{15}$N lines. Higher-resolution observations of this region, with ALMA for example, would be useful in this regard, as well as for disentangling and successfully fitting the potentially multiple gas components in feature SW/S2, and for more precisely modeling the radiative excitation occurring in this region. 
		
		\subsection{CND Chemistry}
		\label{env}
		Recent measurements suggest that the central 300 parsecs of the Galaxy are subject to an elevated cosmic ray ionization rate, ranging from $\zeta \sim 10^{-16}-10^{-14}$ s$^{-1}$ \citep{Oka05,vanderTak06,Goto08,Goto11,Goto13} up to $10^{-13}$ s$^{-1}$ \citep{YZ07, YZ13b,YZ13c}. For comparison, the typical interstellar cosmic ray ionization rate $\zeta_0$ is estimated to be $3\times10^{-17}$  s$^{-1}$, although recent observations by \citealt{Indriolo12} suggest that it may be an order of magnitude higher. A high flux of cosmic rays is suggested to be responsible for heating the molecular gas in the Galactic center, particularly for elevating the observed gas temperatures \citep[50-200 K, e.g.;][]{Gusten85,Huttem93b} above the observed dust temperatures \citep[15-30 K, e.g.;][]{Mezger86,Molinari11} in this region \citep{Gusten81,Ao13,Clark13}. A predicted result of such a high ionization rate is a high fractional ionization of the molecular gas \citep{Papa10,Ao13,YZ13b}: molecular ions such as \hco\, should be more abundant, and this should be reflected in a lowered [HCN]/[\hco] abundance ratio \citep{Meijer06}. A high flux of X-rays could also alter the gas chemistry in this region. As the CND gas is in close proximity to the central supermassive black hole which may have recently (within the past few hundred years) undergone an outburst several orders of magnitude stronger than its typical flares \citep{Koyama96,Murakami00,Inui09,Ponti10,Clavel13}, it is also possible that molecular abundances in this environment have been affected by an enhanced X-ray flux. 
		
		The ratio of observed HCN and \hco\, line intensities has been suggested to be a useful diagnostic for identifying gas in both X-ray dominated (XDR) and cosmic-ray dominated (CRDR) environments \citep{Meijer06, Meijer07}. We investigate the possibility that gas in the CND is subject to either a XDR or CRDR environment by comparing the observed ratio of the HCN and \hco\, $J$ = 4-3 transitions in the CND to those derived in the XDR and CRDR models of \cite{Meijer06, Meijer07}. The observed ratios of HCN 4-3 and \hco\, 4-3 in the CND are between 1.5 and 2.0 (and the ratios of the H$^{13}$CN and \hcoiso\, isotopologues of the same transition, which are free of opacity effects, are even higher, ranging from 2.5 to 5). The predicted HCN 4-3 to  \hco\, 4-3 ratio for a CRDR with a cosmic ray ionization rate of $5\times10^{-15}$ s$^{-1}$ and a density of $10^5$ \cm\, is 0.85 \citep{Meijer06}. Predicted ratios for HCN 4-3 to  \hco\, 4-3 in an XDR with densities higher than $10^5$ \cm are less than 0.4. As HCN 4-3 is significantly stronger than \hco\, 4-3 in the CND, the environment of this gas is not consistent with predictions for molecular gas in a CRDR (at least for densities $\sim$ a few $10^5$ \cm), or an XDR of any density. These observational constraints indicate either that cosmic ray ionization rates in the CMZ are lower than a few times $10^{-15}$ s$^{-1}$, or that the cosmic ray ionization rate varies strongly throughout the CMZ.  

		These observations then suggest that neither X-rays nor cosmic rays are responsible for heating the CND gas, consistent with recent findings by \cite{Goic13} for hot gas interior to the CND, and the findings of \cite{RF04} for the CMZ as a whole. Another possible heating mechanism for the gas is photoelectric heating in photodissociation regions (PDRs):  we find that the measured [HCN]/[\hco] ratios are consistent with the model predictions of \cite{Meijer07} for a relatively high-density PDR ($10^5-10^6$ \cm, similar to densities derived by our excitation analysis). However, PDR heating should be most effective at the surface of the CND surface, and may not be sufficient to explain temperatures of $\gtrsim 200$ K found for the dense CND gas \citep{RT12}. 
				
Mechanical processes could also provide a mechanism to heat the gas in the dense and UV-shielded CND interior consistent with the observed abundances of HCN and \hco.  Where the gas in clump interiors is heated to temperatures $\gtrsim$100 K, HNC will be efficiently converted to HCN \citep{Schilke92,Talbi96}, enhancing the latter's abundance.  \cite{Loenen08} show, for somewhat lower densities than those likely to exist in the CND (n = $10^{4.5}$ \cm), that including the effects of mechanical heating increases the relative abundance of HCN to \hco\, by a factor of 2-3, consistent with our observed HCN/\hco\, ratios. A likely driver of mechanical heating in the CND gas is the cascade of turbulent energy (evident in the large line widths observed in the CND) to smaller size scales where it is dissipated. This turbulence could be injected by the accretion of material onto the CND \citep[e.g.,][]{AB11,Liu12}, or possibly by the nearby Sgr A East supernova remnant \citep{Rock05}, though the interaction between the CND and Sgr A East has not been clearly demonstrated \citep{Lee08,Sjou08}. If the enhanced HCN abundance in the CND is due to mechanical heating in the dense clump interiors, the models of \cite{Loenen08} predict a relatively low HNC/HCN ratio, which can be tested with future observations.
			
\section{Summary}
In this paper, we present observations of multiple lines of HCN and \hco, including the $^{13}$C isotopologues, toward four positions in the circumnuclear disk (CND) of molecular gas and dust in the central parsecs of the Galaxy. We use the measured main beam brightness temperature of each line over the majority of the line profiles (excluding velocity ranges where there is contamination from other nearby clouds) to constrain the H$_2$ volume densities, kinetic temperatures, molecular column densities, and areal filling factors of dense gas in the CND. We also model the physical conditions over limited velocity ranges corresponding to single clumps identified in interferometric studies, in order to determine whether the conditions in individual clumps deviate from the average physical conditions in the CND gas. Our main findings are summarized below: 
 
\begin{enumerate}
\item Using the RADEX radiative transfer code to fit the majority of the line profiles from HCN and \hco, and assuming purely collisional excitation, we find typical densities of log[n(\cm)] $\sim 5.3 - 6.5$, and typical temperatures of T $\sim$ $100-400$ K for the dense gas in the CND, although in some cases the highest temperatures present are unconstrained by these fits. Fitting to limited velocity ranges in the profile corresponding to individual clumps, we derive slightly lower temperatures and slightly higher densities. This suggests conditions in individual clumps may deviate from the average physical conditions in the CND gas. However, from our RADEX models, we find only one feature (S1) which could have a density sufficiently high to be tidally stable; the bulk of the gas in the CND is not stable against tidal shearing. 

\item The average HCN and \hco\, densities for the CND gas indicated by the RADEX fits are consistent within the uncertainties with results from other tracers including CO and dust for all positions except South-1 (for which we find a significantly higher density). In general, the best-fit HCN densities tend to be slightly higher than inferred by these other excitation analyses. 

\item We also detect vibrationally-excited HCN for the first time in the CND. We observe the $J$=4-3 $v_2=1$ line toward the southern emission peak of the CND, at a velocity of -20 \kms\hspace{-0.1cm}, consistent with the SW/S2 clump. We model the excitation of this line by including radiative excitation of HCN due to hot dust in the CND, and find that the observed brightness temperature of this line requires dust temperatures in this region to be $> 125-150$ K. If such hot dust is present in other clumps along the inner edge of the CND, this would have the effect of lowering the densities we find with HCN. Dust temperatures of T$\sim$150 K are sufficient to bring the densities we derive for the southern emission peak into agreement with those found using other tracers, such as CO and dust, and would then make it unlikely that {\em any} of the gas in the CND is tidally stable. 

\end{enumerate}

\section{Acknowledgements}
The authors would like to thank Ryan Lau for sharing his maps of the CND dust opacity and temperature in advance of publication. We would also like to thank Peter Schilke for very useful comments and discussion during the preparation of this paper. We are also grateful to the anonymous referee for their insightful comments which improved the discussion in this paper.
This research has made use of the NASA/ IPAC Infrared Science Archive, which is operated by the Jet Propulsion Laboratory, California Institute of Technology, under contract with the National Aeronautics and Space Administration.

\bibliographystyle{apj}
\bibliography{APEX.bib}

\clearpage

\begin{table}[p!]
\caption{Properties of the Observed Transitions} 
\centering
\begin{tabular}{llcccc}
\\[0.5ex]
\hline\hline
& & & & & \\
{\bfseries Molecule}&{\bfseries Transition}&{\bfseries Frequency}&{\bfseries Upper State Energy}&{\bfseries Critical Density\footnotemark[1]}&{\bfseries APEX beam}\\ [0.5ex]
& $J$--$J$-$1$, $v_2$ & [GHz] & [K] & [\cm] & FWHM\\
\hline
HCN & 3--2      &  265.88618 & 25.5 & 5.2$\times10^7$ & 23.6$''$ \\
	& 4--3       &  354.50548 & 42.5 & 1.1$\times10^8$ & 17.7$''$ \\
	& 4--3, $1f$ & 356.25561 & 1067.1 &  \footnotemark[2]3.1$\times10^{11}$ & 17.6$''$ \\
	& 8--7       & 708.87721 & 153.1 & 8.7$\times10^8$ & 8.9$''$ \\
\hline
HC$^{13}$N & 3--2 & 259.01182 & 24.9 & 4.8 $\times10^7$ & 24.3$''$ \\
		      & 4--3 & 345.33976 & 41.4& 1.1$\times10^8$ & 18.2$''$ \\
\hline		    
\hco & 3--2      & 267.55753 & 25.7 &  3.5$\times10^6$ & 23.5$''$ \\
	& 4--3        & 356.73413 & 42.8 & 8.2$\times10^6$ & 17.6$''$ \\
	& 9--8	& 802.45822 & 192.6 & 1.0$\times10^8$ & 7.8$''$\\
\hline
\hcoiso & 3--2 & 260.25534 & 25.0 & 3.1$\times10^6$ & 24.2$''$\\
	     & 4--3 & 346.99834 & 41.6 & 7.3$\times10^6$ & 18.1$''$\\
\hline
\footnotetext[1]{Assuming a kinetic temperature of 200 K}
\footnotetext[2]{This is actually the critical density required to populate the $J$=4, $v_2$=1 level, using a collision rate estimated from CO$_2$ at $\sim$ 200 K \citep{Ziurys} }
\end{tabular}
\label{lines}
\end{table}
\clearpage

\begin{table}[p!]
\caption{\vspace{0.2cm}{ \large Clumps}} 
\centering
\begin{small}
\begin{tabular}{l c c c c}
\hline
{\bfseries Name} & {\bfseries Offset (RA, Dec)} & {\bfseries Velocity} & {\bfseries Alternate} & {\bfseries Alternate} \\
& & &  {\bfseries ID \#1}\footnotemark[1] & {\bfseries ID \#2}\footnotemark[2] \\
\hline
{\bfseries N} & (+25\arcsec,+40\arcsec)& 90 to 110 \kms & D & A\\
{\bfseries S1} & ( --20\arcsec, --30\arcsec  ) & -115  to -95 \kms & O & Q\\
{\bfseries S2} & ( --20\arcsec, --30\arcsec  )& -25 to -5 \kms & -- & --\\
{\bfseries SW} & ( --30\arcsec, --20\arcsec ) & -25 to -5 \kms & P & N\\
{\bfseries W} & ( --20\arcsec, +0\arcsec )& 35 to 55 \kms & T & K \\
\footnotetext[1]{ \cite{Chris05}}
\footnotetext[2]{ \cite{MCHH09}}
\end{tabular}
\end{small}
\label{Clumps}
\end{table}
\clearpage

\begin{table}[p!]
\caption{\vspace{0.2cm}{ \large Integrated Brightness Temperatures}} 
\centering
\begin{small}
\begin{tabular}{c c c c c|c c|c c c|c c }
\multicolumn{11}{c}{ \bfseries $\int$ T$_{MB}$ d$v$ (K km s$^{-1}$ ) } \\\\
 \hline
&  {\bfseries HCN} & & & &  {\bfseries H$^{13}$CN}  & & {\bfseries HCO$^+$} & & & {\bfseries H$^{13}$CO$^+$} &  \\
& 3--2 & 4--3 & 8--7 & &  3--2 & 4--3 & 3--2 & 4--3 & 9--8 & 3--2 & 4--3 \\
\hline
\multicolumn{12}{c}{\bfseries Integrated over the majority of the line profile} \\
\hline
\multicolumn{4}{l}{{\bfseries North} {\bfseries \emph {v = 20 to 120 \kms}}} & & & & & & & &\\
& 198$\pm$20 & 137$\pm$14 &  27.3$\pm$ 5.5 & & 17.9$\pm$1.8 & 10.8$\pm$1.1 & 127$\pm$13 &  93.2$\pm$ 9.3 &   2.8$\pm$ 0.6 &  5.6$\pm$0.6 &  4.1$\pm$0.4 \\
\multicolumn{4}{l}{{\bfseries South} {\bfseries \emph {v = -120 to -20 \kms}}} & & & & & & & &\\
& 340$\pm$34 & 285$\pm$29 &  62$\pm$12 & & 33.3$\pm$3.3 & 28.1$\pm$2.8 & 173$\pm$17 & 140.4$\pm$14 &  14.2$\pm$ 2.9 &  7.4$\pm$0.7 &  6.1$\pm$0.6 \\
\multicolumn{4}{l}{{\bfseries South} {\bfseries \emph{v = -50 to 5 \kms}}} & & & & & & & & \\
& 152$\pm$15& 117$\pm$12 &  25.3$\pm$ 5.1 & & 29.1$\pm$2.9 & 17.1$\pm$1.7 &  81.7$\pm$ 8.2 &  67.8$\pm$ 6.8 &   7.7$\pm$ 1.6 &  7.5$\pm$0.8 &  4.4$\pm$0.4 \\
\multicolumn{4}{l}{{\bfseries Southwest} {\bfseries \emph{ v = -50 to 5 \kms}}}& & & & & & & &\\
& 159$\pm$16 & 137$\pm$14 &  36.0$\pm$ 7.2 & & 22.5$\pm$2.3 & 13.2$\pm$1.3 &  82.1$\pm$ 8.2 &  86.0$\pm$ 8.6 &   10.1$\pm$ 2.0 &  7.5$\pm$0.8 &  4.4$\pm$0.4 \\
\multicolumn{4}{l}{{\bfseries West,} {\bfseries \emph{v = 0 to 100 \kms}}} & & & & & & & &\\
& 194$\pm$19 & 124$\pm$12 &  20.4$\pm$ 4.1 & & 17.1$\pm$1.7 &  9.8$\pm$1.0 & 111$\pm$11 &  83.7$\pm$ 8.4 &   5.8$\pm$ 1.2 &  3.2$\pm$0.3 &  4.4$\pm$0.4 \\
\hline
\multicolumn{12}{c}{\bfseries Integrated over the velocity ranges of individual features (reported in Table \ref{Clumps})} \\
\hline
 {\bfseries  N} & 49.0$\pm$4.9 & 35.0$\pm$3.5 &  7.5$\pm$1.5 &  & 5.6$\pm$0.6 &  4.3$\pm$0.4 & 33.8$\pm$3.4 & 23.5$\pm$2.4 &  0.7$\pm$0.1 &  1.7$\pm$0.2 &  1.7$\pm$0.2 \\
 {\bfseries S1} & 57.2$\pm$5.7 & 49.9$\pm$5.0 & 11.9$\pm$2.4 &  &4.4$\pm$0.4 &  5.8$\pm$0.6 & 29.5$\pm$3.0 & 24.3$\pm$2.4 &  3.4$\pm$0.7 &  0.6$\pm$0.1 &  1.2$\pm$0.1 \\
 {\bfseries S2} & 38.0$\pm$3.8 & 35.8$\pm$3.6 &  8.7$\pm$1.7 & &12.2$\pm$1.2 &  7.1$\pm$0.7 & 21.6$\pm$2.2 & 23.4$\pm$2.3 &  3.0$\pm$0.6 &  3.8$\pm$0.4 &  1.9$\pm$0.2 \\
 {\bfseries SW} & 62.6$\pm$6.3 & 68.8$\pm$6.9 & 18.3$\pm$3.7 & &13.1$\pm$1.3 &  8.2$\pm$0.8 & & & & & \\
 {\bfseries  W} & 57.9$\pm$5.8 & 37.0$\pm$3.7 &  7.6$\pm$1.5 &  &5.2$\pm$0.5 &  3.1$\pm$0.3 & 35.9$\pm$3.6 & 26.1$\pm$2.6 &  4.0$\pm$0.9 &  1.3$\pm$0.1 &  1.3$\pm$0.1 \\
\end{tabular}
\end{small}
\label{Intensity}
\end{table}
\clearpage

\begin{table}[p!]
\caption{\vspace{0.2cm}{ \large Results of RADEX Excitation Analysis}} 
\centering
\begin{small}
\begin{tabular}{lccccccccccc}
\hline\hline
&   {\bfseries min. $\chi^2$} & {\bfseries T$_{\textrm{kin}}$} & {\bfseries log [n$_{\textrm{H}_2}$]} & {\bfseries log [N$_{\textrm{Mol}}$]} & {\bfseries $f$} & {\bfseries Radius}&{\bfseries Mass}&{\bfseries $\tau_{ 3-2}$}&{\bfseries $\tau_{ 4-3}$}\\
&  & [K] & [\cm] &  [cm$^{-2}$] & & [pc] & [M$_\odot$] &  & \\
\hline
& \multicolumn{7}{l}{{\bfseries HCN fits} (majority of the line profile)} \\
\hline
\vspace{0.1cm}
\rule{0pt}{4ex}{\bfseries North} (v=20,120) & 2.0 & 270$_{-132}^{+319}$ & 5.6$_{-0.6}^{+0.6}$ & 16.2$_{-0.2}^{+0.3}$ & 0.04$_{-0.02}^{+0.01}$ &&& 5.0$_{-2.3}^{+5.7}$ & 7.1$_{-3.2}^{+7.4}$ \\
\vspace{0.1cm}
{\bfseries South-1} (v=-120,-20) & 0.8 & 94$_{-33}^{+198}$ & 6.5$_{-0.7}^{+0.5}$ & 16.0$_{-0.1}^{+0.2}$ & 0.07$_{-0.02}^{+0.01}$ &&& 2.1$_{-0.4}^{+1.9}$ & 3.1$_{-0.6}^{+2.8}$ \\
\vspace{0.1cm}
{\bfseries South-2} (v=-50,5) & 1.3 & 94$_{-22}^{+11}$ & 5.9$_{-0.2}^{+0.4}$ & 16.4$_{-0.3}^{+0.1}$ & 0.05$_{-0.01}^{+0.01}$ &&& 10.3$_{-5.0}^{+5.2}$ & 15.0$_{-7.3}^{+6.3}$ \\
\vspace{0.1cm}
{\bfseries Southwest} (v=-50,5) & 0.5 & 171$_{-66}^{+88}$ & 5.9$_{-0.2}^{+0.4}$ & 16.1$_{-0.2}^{+0.2}$ & 0.06$_{-0.01}^{+0.01}$ &&& 4.6$_{-1.7}^{+4.0}$ & 7.0$_{-2.4}^{+5.7}$ \\
\vspace{0.1cm}
{\bfseries West} (v=0,100) & 3.1 & 204$_{-88}^{+220}$ & 5.6$_{-0.5}^{+0.6}$ & 16.3$_{-0.3}^{+0.2}$ & 0.04$_{-0.01}^{+0.01}$ &&&9.8$_{-7.2}^{+1.7}$ & 13.8$_{-10.0}^{+1.5}$ \\
\hline
&\multicolumn{7}{l}{{\bfseries HCN fits} (narrow features)} \\
\hline
\vspace{0.1cm}
\rule{0pt}{4ex}{\bfseries N} (v=90,110) & 3.0 & 94$_{-22}^{+77}$ & 6.3$_{-0.8}^{+0.4}$ & 15.5$_{-0.2}^{+0.9}$ & 0.04$_{-0.02}^{+0.01}$ &0.10$_{-0.03}^{+0.01}$&480$_{-450}^{+1100}$& 3.8$_{-1.4}^{+3.3}$ & 5.6$_{-2.0}^{+17.3}$ \\
\vspace{0.1cm}
{\bfseries S1} (v=-115,-95) & 1.7 & 61$_{-11}^{+77}$ & 7.2$_{-0.7}^{+0.4}$ & 15.1$_{-0.1}^{+0.1}$ & 0.09$_{-0.02}^{+0.02}$ &0.15$_{-0.02}^{+0.01}$&13000$_{-11000}^{+34000}$& 1.5$_{-0.3}^{+0.4}$ & 2.1$_{-0.2}^{+0.7}$ \\
\vspace{0.1cm}
{\bfseries S2} (v=-25,-5) & 0.8 & 72$_{-11}^{+11}$ & 5.7$_{-0.2}^{+0.6}$ & 16.4$_{-0.4}^{+0.1}$ & 0.04$_{-0.01}^{+0.01}$ &0.10$_{-0.02}^{+0.01}$&120$_{-40}^{+420}$& 26.4$_{-17.2}^{+11.1}$ & 37.1$_{-23.8}^{+14.2}$ \\
\vspace{0.1cm}
{\bfseries SW} (v=-25,-5) & 3.1 & 116$_{-33}^{+11}$ & 6.0$_{-0.5}^{+0.4}$ & 15.9$_{-0.2}^{+0.5}$ & 0.04$_{-0.01}^{+0.01}$ & 0.10$_{-0.02}^{+0.01}$&240$_{-160}^{+560}$& 6.8$_{-2.6}^{+25.4}$ & 10.4$_{-4.0}^{+21.2}$ \\
\vspace{0.1cm}
{\bfseries W} (v=35,55) & 3.7 & 237$_{-110}^{+286}$ & 5.6$_{-0.6}^{+0.5}$ & 15.5$_{-0.2}^{+0.3}$ & 0.06$_{-0.02}^{+0.02}$ &0.12$_{-0.02}^{+0.02}$&170$_{-110}^{+660}$& 4.8$_{-2.0}^{+5.9}$ & 6.9$_{-2.7}^{+10.9}$ \\
 \hline
 &\multicolumn{7}{l}{{\bfseries \hco\, fits} (majority of the line profile)} \\
\hline
\vspace{0.1cm}
\rule{0pt}{4ex}{\bfseries North} (v=20,120) & 0.2 & 94$_{-22}^{+55}$ & 5.9$_{-0.4}^{+0.2}$ & 14.4$_{-0.7}^{+0.5}$ & 0.22$_{-0.14}^{+0.78}$ &&& 0.3$_{-0.2}^{+0.7}$ & 0.4$_{-0.3}^{+0.7}$ \\
\vspace{0.1cm}
{\bfseries South-1} (v=-120,-20) & 0.2 &  \footnotemark[1]457$_{-231}^{+132}$ & 5.6$_{-0.2}^{+0.2}$ & 14.4$_{-0.6}^{+0.4}$ & 0.29$_{-0.18}^{+0.71}$ &&& 0.2$_{-0.2}^{+0.4}$ & 0.3$_{-0.2}^{+0.4}$ \\
\vspace{0.1cm}
{\bfseries South-2} (v=-50,5) & 9.7 & & & & & & \\
\vspace{0.1cm}
{\bfseries Southwest} (v=-50,5) & 7.3 & & & & & & \\
\vspace{0.1cm}
{\bfseries West} (v=0,100) & 7.4 & & & & & & \\
\hline
 &\multicolumn{7}{l}{{\bfseries \hco\, fits} (narrow features)} \\
\hline
\vspace{0.1cm}
\rule{0pt}{4ex}{\bfseries N} (v=90,110) & 4.5 & 61$_{-11}^{+22}$ & 6.2$_{-0.3}^{+0.2}$ & 14.5$_{-0.3}^{+0.2}$ & 0.07$_{-0.02}^{+0.06}$ &0.13$_{-0.02}^{+0.06}$&840$_{-590}^{+3300}$& 0.9$_{-0.2}^{+0.3}$ & 1.3$_{-0.3}^{+0.3}$ \\
\vspace{0.1cm}
{\bfseries S1} (v=-115,-95) & 3.5 & \footnotemark[1]336$_{-165}^{+253}$ & 5.9$_{-0.2}^{+0.2}$ & 13.1$_{-0.0}^{+0.7}$ & 1.00$_{-0.80}^{+0.00}$ &&& 0.0$_{-0.0}^{+0.0}$ & 0.1$_{-0.0}^{+0.0}$ \\
\vspace{0.1cm}
{\bfseries S2} (v=-25,-5) & 19.8 & & & & & & \\
\vspace{0.1cm}
{\bfseries SW} (v=-25,-5) & 23.5 & & & & & & \\
\vspace{0.1cm}
{\bfseries W} (v=35,55) & 4.0 & 589$_{-352}^{+0}$ & 5.6$_{-0.1}^{+0.2}$ & 13.3$_{-0.2}^{+0.9}$ & 0.69$_{-0.59}^{+0.31}$ &&& 0.1$_{-0.0}^{+0.3}$ & 0.1$_{-0.0}^{+0.4}$ \\
\hline
\footnotetext[1]{Upper bound on the temperature is unconstrained}
\end{tabular}
\end{small}
\label{Results}
\end{table}
\clearpage

\end{document}